\documentclass[a4paper, amsfonts, amssymb, amsmath, reprint, showkeys, nofootinbib, twoside]{revtex4-2}
\usepackage[english]{babel}
\usepackage[utf8]{inputenc}
\usepackage[colorinlistoftodos, color=green!40, prependcaption]{todonotes}

\usepackage{graphicx}

\usepackage[utf8]{inputenc}

\usepackage{amsmath}
\usepackage{mathtools}
\usepackage{amssymb}
\usepackage{cases}

\usepackage{braket}

\usepackage{amsthm}
\usepackage{mathtools}
\usepackage{physics}
\usepackage{xcolor}
\usepackage{graphicx}
\usepackage[left=23mm,right=13mm,top=35mm,columnsep=15pt]{geometry} 
\usepackage{adjustbox}
\usepackage{placeins}
\usepackage[T1]{fontenc}
\usepackage{lipsum}
\usepackage{csquotes}

\usepackage{amsmath}
\usepackage{amssymb}

\usepackage{braket}

\usepackage{natbib}

\usepackage[pdftex, pdftitle={Article}, pdfauthor={Author}]{hyperref} 
\begin{document}
\title{Modeling and Experimental Validation of the Intrinsic SNR in Spin Qubit Gate-Based Readout and Its Impacts on Readout Electronics}

\author{Bagas Prabowo\textsuperscript{1,2}}
    \email[]{B.Prabowo@tudelft.nl}
    \author{Jurgen Dijkema\textsuperscript{2,3}}
    \author{Xiao Xue\textsuperscript{2,3}} 
    \author{\mbox{Fabio Sebastiano}\textsuperscript{1,2}}   
    \author{Lieven M. K. Vandersypen\textsuperscript{2,3}}   
    \author{Masoud Babaie\textsuperscript{1,2}}   
    \affiliation{\textsuperscript{1}Department of Quantum and Computer Engineering, Delft University of Technology, 2600 Delft, GA, The Netherlands}
    \affiliation{\textsuperscript{2}QuTech, Delft University of Technology, 2600 Delft, GA, The Netherlands}
    \affiliation{\textsuperscript{3}Kavli Institute of Nanoscience, Delft University of Technology, 2600 Delft, GA, The Netherlands}
\date{13-12-2023} 

\begin{abstract}
In semiconductor spin quantum bits (qubits), the radio-frequency (RF) gate-based readout is a promising solution for future large-scale integration, as it allows for a fast, frequency-multiplexed readout architecture, enabling multiple qubits to be read out simultaneously. This paper introduces a theoretical framework to evaluate the effect of various parameters, such as the readout probe power, readout chain's noise performance, and integration time on the intrinsic readout signal-to-noise ratio (SNR), and thus readout fidelity of RF gate-based readout systems. By analyzing the underlying physics of spin qubits during readout, this work proposes a qubit readout model that takes into account the qubit's quantum mechanical properties, providing a way to evaluate the trade-offs among the aforementioned parameters. The validity of the proposed model is evaluated by comparing the simulation and experimental results. The proposed analytical approach, the developed model, and the experimental results enable designers to optimize the entire readout chain effectively, thus leading to a faster, lower-power readout system with integrated cryogenic electronics.
\end{abstract}


\maketitle

\section{Introduction}

Performing practical, fault-tolerant quantum algorithms with quantum error correction (QEC) will require thousands of quantum bits (qubits) in a quantum computer to be manipulated and read out simultaneously \cite{fowlerSurfaceCodesPractical2012}. In the case of a semiconductor spin qubit quantum computer (QC), the QEC cycle must be able to read, decode the error, and apply the correction to the data qubits far faster than the qubit's dephasing time (T\textsubscript{2}\textsuperscript{*}) \cite{overwater_neural-network_2022}. Owing to the short T\textsubscript{2}\textsuperscript{*} of $\sim$20\,$\mu$s \cite{struck_low-frequency_2020,yoneda_quantum-dot_2018,xue_quantum_2022}, it is essential to reduce the readout, decode, and gate operation time to be below the sub-microsecond range for the QEC cycle time to be significantly faster than the dephasing time. 

In the case of reducing the readout time of the quantum processor, the choice in the readout technique plays a crucial role. With the requirement of reading thousands of qubits in the future, the choice of the readout architecture must be fast and scalable. One of the most promising readout techniques pursued currently for semiconductor spin qubits is the gate-based readout technique, as it allows the readout architecture to adopt frequency division multiplex access (FDMA) feature to enable the simultaneous readout of multiple qubits by a single receiver (RX) \cite{Li_Petit_2018, ruffino_cryo-cmos_2022, hornibrook_frequency_2014}.  


 \begin{figure}
    \centering
    \includegraphics[width=0.40\textwidth]{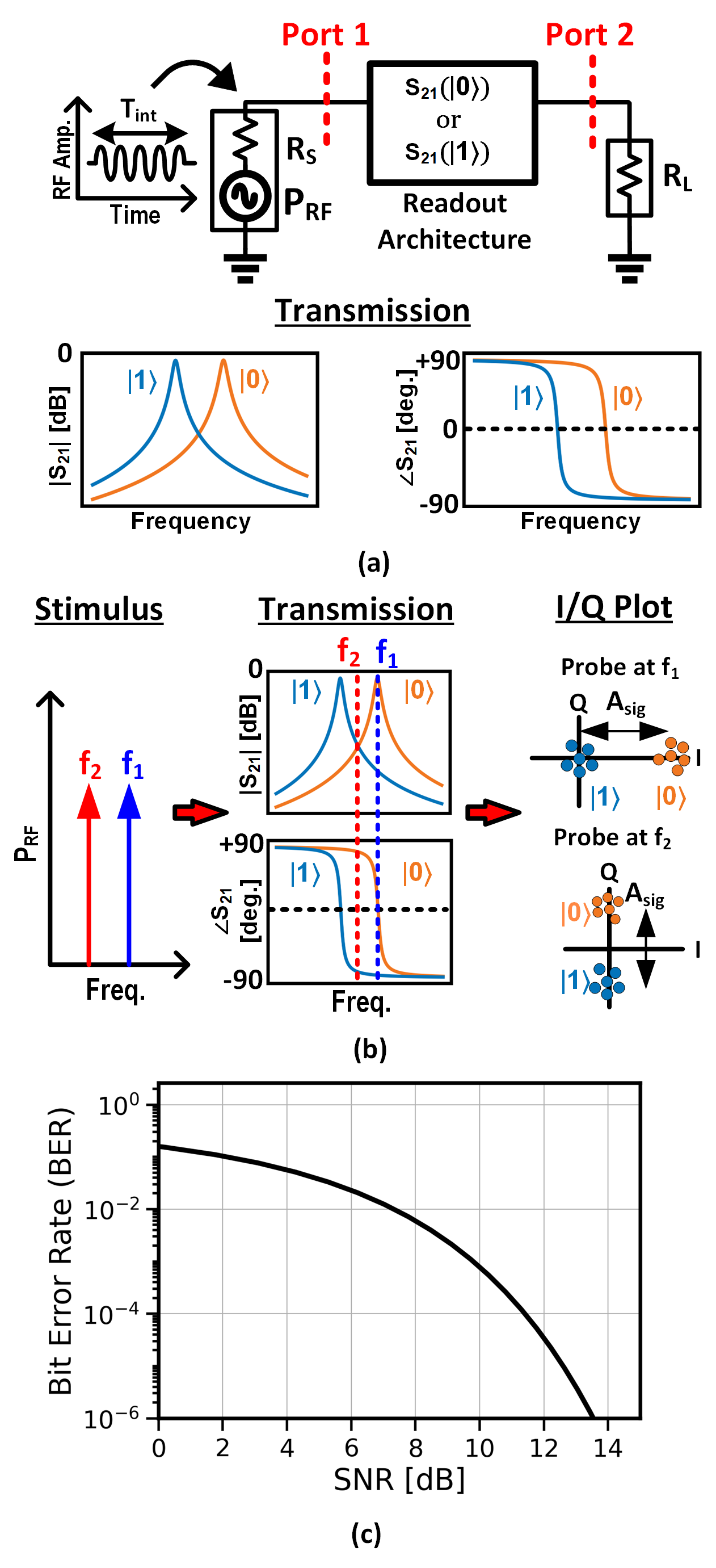}
    \caption{(a) A prototype of a readout architecture; (b) Corresponding probe tones, transmission responses of the sample, and constellation diagrams when the system is probed at $f_1$ and $f_2$; (c) BER versus SNR plot. \label{fig:architecture_proposal}}
\end{figure}

A simplified model of a gate-based readout architecture is illustrated in Fig.\,\ref{fig:architecture_proposal}(a), where port-1 and port-2 indicate the input and output ports of the readout system, respectively. In principle, the readout architecture has a state-dependent transmission (or reflection) behavior dependent on the qubit's state. To read the qubit's state, the system is probed with a radio-frequency (RF) tone from port-1 with a certain time duration, $T_{int}$. Depending on the frequency of the stimulus tone, the readout tone observed at port-2 exhibits a change in amplitude or phase, providing information about the qubit's state. As an example, Fig.\,\ref{fig:architecture_proposal}(b) illustrates the readout system's response when probed at $f_1$ and $f_2$. When the system is probed at the resonance frequency of the $\ket{0}$ state ($f_1$), the readout signal observed at port-2 resembles a binary amplitude shift keying signal (BASK), due to the attenuated $\ket{1}$ response. On the other hand, when the system is probed at a frequency precisely between the $\ket{0}$ or $\ket{1}$ responses ($f_2$), the output signal at port-2 resembles a binary phase shift keying signal (BPSK), where the $\ket{0}$ and $\ket{1}$ response differs by 180 degrees. Regardless of how the output signal is modulated, the readout signal can be denoted as $A_{sig}$ in the constellation diagram, which describes the separation between the two states' responses. 

In general, the readout signal (i.e., $A_{sig}$) must be large enough to accommodate the thermal noise contributed by the environment and the additional noise of the subsequent readout electronics, which is responsible for amplifying and down-converting the readout signal. To quantify the effect of the noise on the readout signal, the readout signal-to-noise ratio (SNR) must be calculated, which depends on both the power of the readout signal ($P_{Sig}$) and the total noise observed at the input of the readout electronics ($P_{N}$). Mathematically, $P_{Sig}$ is related to $A_{sig}$ by
\begin{align}
    P_{Sig} = \frac{|A_{sig}|^2}{R_L} = P_{RF}\left| \left(S_{21,\ket{0}}(\omega_r) - S_{21,\ket{1}}(\omega_r)\right)\right|^2 ,\label{eq:P_sig_intro}
\end{align}
where $P_{RF}$ is the power delivered by the RF source to the input port of the system (i.e., port-1), $\omega_r$ is the readout angular frequency and $\left| \left(S_{21,\ket{0}}(\omega_r) - S_{21,\ket{1}}(\omega_r)\right)\right|^2$ is the \textit{state-separation factor}, which depends on the $S_{21}$ behavior of the system when it is in the $\ket{0}$ or $\ket{1}$ state. The noise power, $P_N$, on the other hand, can be expressed as $N_0/T_{int}$, where $N_0$ is the noise power spectral density, which includes the noise of the electronics referred to port-2, and $T_{int}$ is the integration time. By defining the SNR as $P_{Sig}/P_{N}$, the readout SNR can be expressed as  
\begin{align}
    SNR = \frac{P_{RF}}{N_0/T_{int}} \times  \left| \left(S_{21,\ket{0}}(\omega_r) - S_{21,\ket{1}}(\omega_r)\right)\right|^2. \label{eq:SNR_theoretical_intro} 
\end{align}
However, the quality of the readout acquisition is commonly characterized using metrics such as the bit error rate (BER) (equivalently, the readout infidelity, $1-F$) rather than the SNR. The relationship between the SNR in (\ref{eq:SNR_theoretical_intro}) with the BER can be described as
\begin{align}
    BER = Q\left(\sqrt{SNR}\right) \label{eq:BER_intro},
\end{align}
where $Q(\cdot)$ is the tail distribution function of the standard normal distribution. Based on (\ref{eq:BER_intro}), the resulting BER for a given SNR performance is shown in Fig.\,\ref{fig:architecture_proposal}(c). In fault-tolerant QC with QEC, a target BER of 10\textsuperscript{-4} is often desired, which corresponds to an SNR requirement of 11.5 dB.

The target SNR of 11.5 dB can be achieved in numerous ways, based on (\ref{eq:SNR_theoretical_intro}). Given a readout system with a certain state-separation factor, one could choose an arbitrary $P_{RF}$, $N_0$, and a $T_{int}$ that complies with the desired SNR. However, in the context of QEC, $T_{int}$ is commonly fixed based on the QEC requirements. Thus, only $P_{RF}$ and $N_0$ are left to be optimized to achieve the required SNR. By maximizing $P_{RF}$, $N_0$ can be made larger while still achieving the desired SNR. The larger tolerable $N_0$ benefits the readout electronics by allowing them to be designed at a lower power. With the limited power budget available at the cryogenic level of the dilution refrigerator, this approach enables better integration of cryogenic electronics suitable for future large-scale integration.

Experimental observations, however, have shown that increasing $P_{RF}$ does not necessarily improve the readout SNR \cite{zhengRapidGatebasedSpin2019,Imtiaz_gatebased_2018, Ibberson_Large_dispersive_2021}. Theoretically, (\ref{eq:SNR_theoretical_intro}) shows a boundless trade-off between $P_{RF}$ and $N_0$, where a large $P_{RF}$ allows $N_0$ to be increased proportionally for a given target SNR, $T_{int}$, and state-separation factor, contradicting experimental observation. Thus, the limit in which $N_0$ can be relaxed as $P_{RF}$ increases remains unclear. In this article, we investigate the validity of (\ref{eq:SNR_theoretical_intro}) by first constructing a qubit model that considers its quantum mechanical behavior during readout. The proposed model is used to simulate the behavior of a readout system to investigate the validity of (\ref{eq:SNR_theoretical_intro}) and the limitations of the trade-offs between $P_{RF}$ and $N_0$. To confirm the behavior shown by our proposed model, the simulated results are compared with experimental results.

The article is organized as follows. Section II first discusses the behavior of the qubits during readout and how they can be modeled. Section III presents the simulation results of the readout system when the proposed qubit model is considered. Section IV compares the simulated results with experimental results, verifying the validity and limitations of the model. Upon experimental verification, Section V discusses the impact of the trade-offs between $P_{RF}$ and $N_0$ on the readout architecture. Section VI concludes the article.

\section{Spin Qubit Readout Theory} \label{sec:Spin Qubit Readout Theory}
\subsection{Double Quantum Dot} %


\begin{figure}
    \centering
    \includegraphics[width=0.48\textwidth]{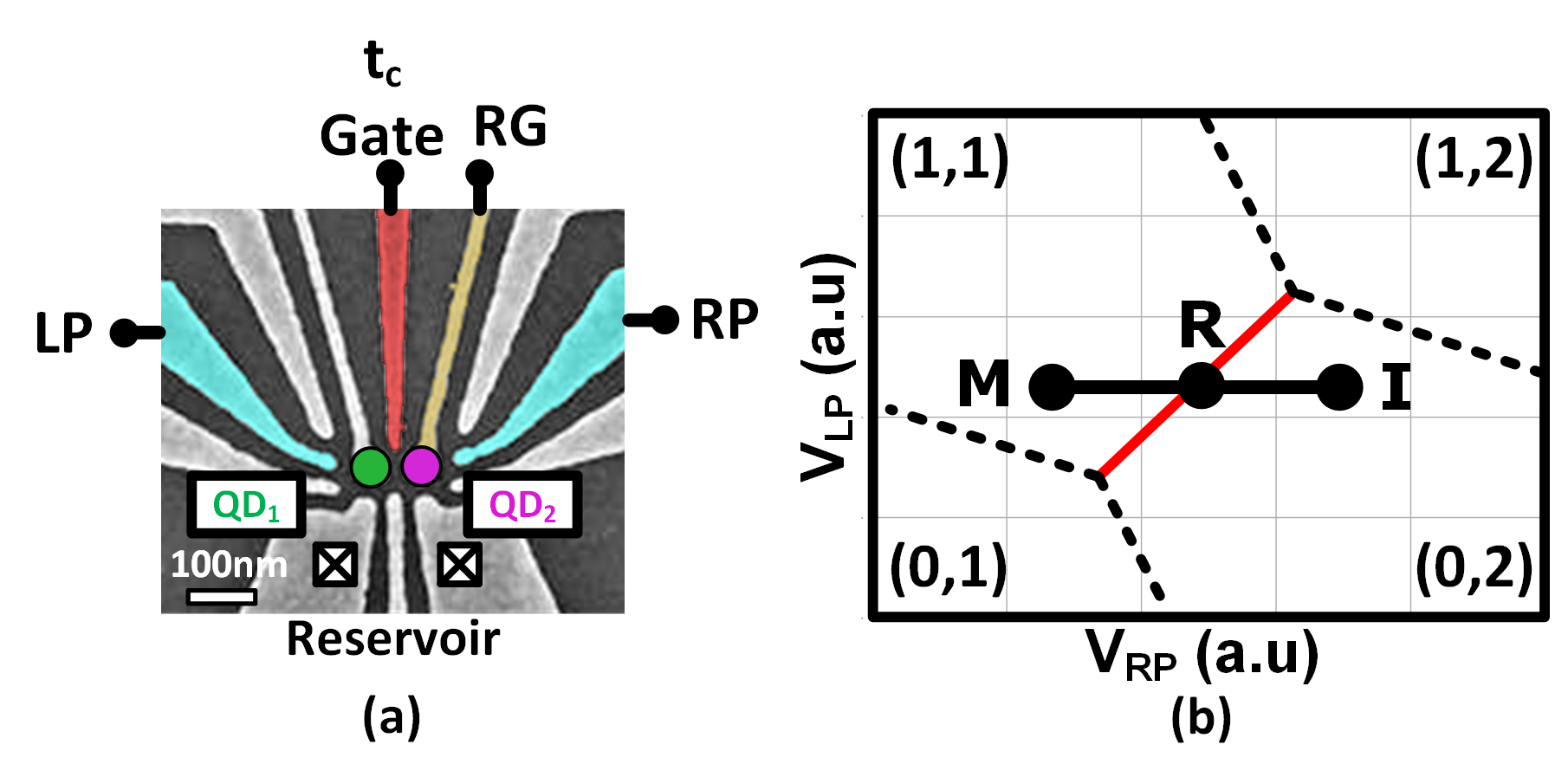} 
    \caption{(a) Colored image of a DQD; (b) Sketch of the charge stability diagram of the DQD. The red solid lines indicate the interdot regime. Initialization, manipulation, and readout regimes are labeled correspondingly. \label{fig:DQD_model}}
\end{figure}

The double quantum dot (DQD) is the basic building block of a semiconductor spin qubit. Similar to a transistor, the DQD is composed of several plunger gates to control the potential landscape of the device to trap electrons in each respective quantum dot (QD), as indicated by the green and purple circles in Fig.\,\ref{fig:DQD_model}(a). By varying the left ($V_{LP}$) and right ($V_{RP}$) plunger gate voltages, individual electrons can be loaded in and out of each QD site from the electron reservoir, as indicated by the crossed boxes in Fig.\,\ref{fig:DQD_model}(a). The tunnel coupling ($t_c$) gate shown in the figure controls the tunnel coupling interaction between the two QDs and plays an essential role in tuning the device for readout.

The DQD is typically characterized by the charge stability diagram, as shown in Fig.\,\ref{fig:DQD_model}(b). The number (N\textsubscript{L},N\textsubscript{R}) indicates the number of electrons in the left and right QD, respectively. The dotted black lines in the diagram indicate the plunger gate voltages at which the electrons from the QD can be loaded or unloaded to the reservoir. For quantum computation with spin qubits, initialization can take place in the (0,2) regime, marked as 'I' in Fig.\,\ref{fig:DQD_model}(b). Once initialized, qubit manipulation can be executed in the (1,1) charge regime at point 'M', where the electron spin in the left QD can be manipulated independently while keeping the right QD unaffected such that it can be used as a reference for readout \cite{xue_quantum_2022}. To read the spin state of the left QD, the DQD is biased at the interdot regime (highlighted in red in Fig.\,\ref{fig:DQD_model}(b)). Depending on the final spin state to which the DQD collapses, the electron may oscillate or stay in each respective dot in response to an RF tone applied at the readout gate (RG), leading to a state-dependent behavior, as discussed in detail below. 


\begin{figure}
    \centering
    \includegraphics[width=0.47\textwidth]{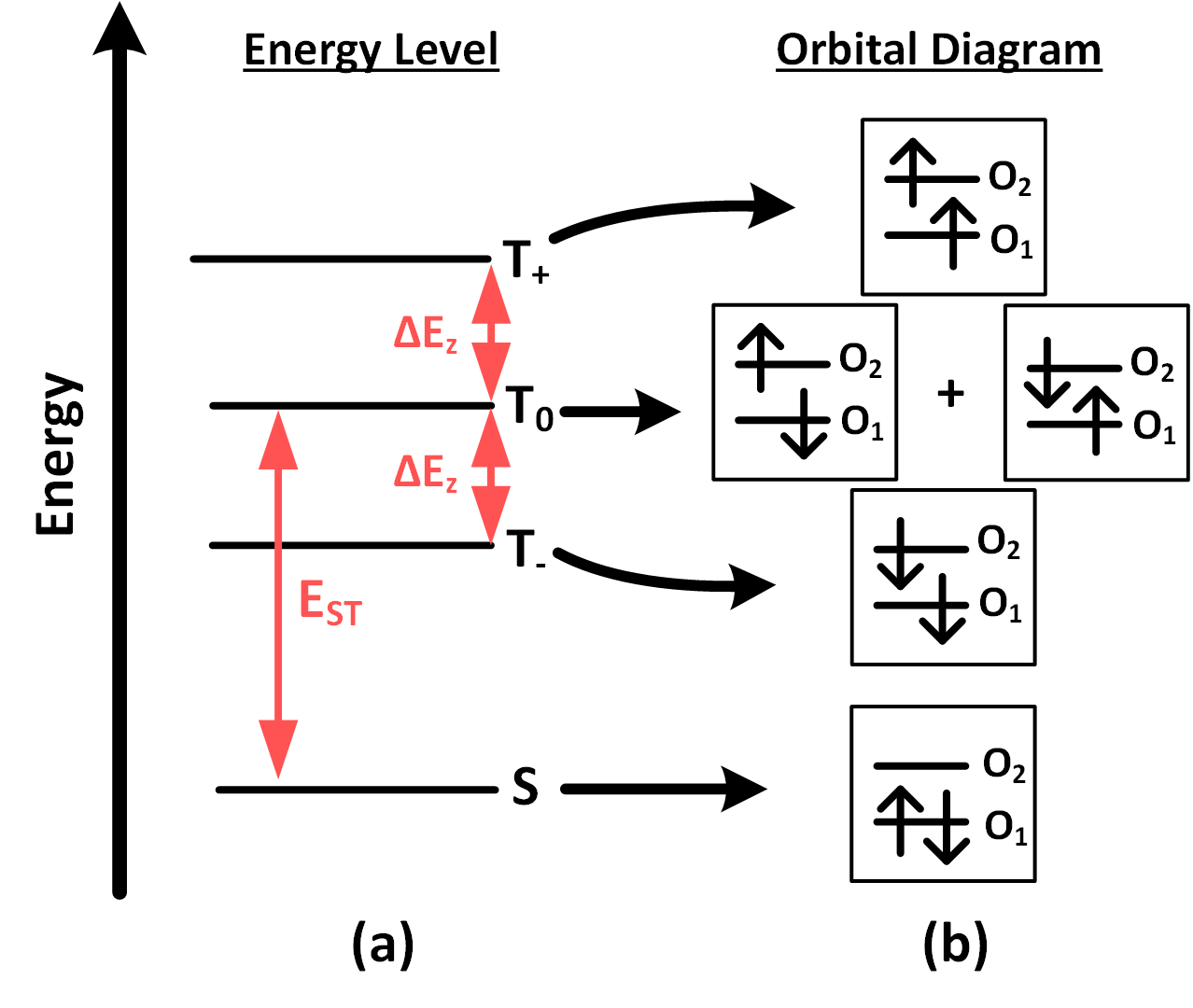} 
    \caption{(a) Energy levels of the singlet and triplet electron configuration; (b) Corresponding electron occupation in the two lowest orbitals. \label{fig:Energy_Orbital_Diagram}}
\end{figure}

The spin states of the DQD and their energy levels in the (1,1)-(0,2) charge regimes must be evaluated to understand how the state-dependent behavior is manifested during readout. Generally, a two-electron system has four possible spin states when subjected to a magnetic field within a QD site. The four possible configurations are namely the singlet (S) and the three triplets (T\textsubscript{-}, T\textsubscript{0}, and T\textsubscript{+}) states, as illustrated in Fig.\,\ref{fig:Energy_Orbital_Diagram}(a). The difference in energy between the T\textsubscript{0} and the T\textsubscript{-} (T\textsubscript{+}) state is denoted as the Zeeman energy ($\Delta$E\textsubscript{z}), while E\textsubscript{ST} denotes the energy difference between the S and T\textsubscript{0}.

Fig.\,\ref{fig:Energy_Orbital_Diagram}(b) depicts the corresponding orbital diagram for the two-electron system for the singlet and triplet states. Only the two lowest orbitals in which electrons can occupy are considered (i.e., O\textsubscript{1} and O\textsubscript{2}). When the DQD is biased in the (0,2) charge regime, these orbitals are spatially confined to only the right QD site, while they spread out across the DQD when biased in the (1,1) charge regime. Due to the Pauli exclusion principle, each orbital can only be occupied by two electrons of opposite spins \cite{hansonSpinsFewelectronQuantum2007}. Consequently, only the singlet state can contain two electrons occupying the same orbital.

\begin{figure}
    \centering
    \includegraphics[width=0.45\textwidth]{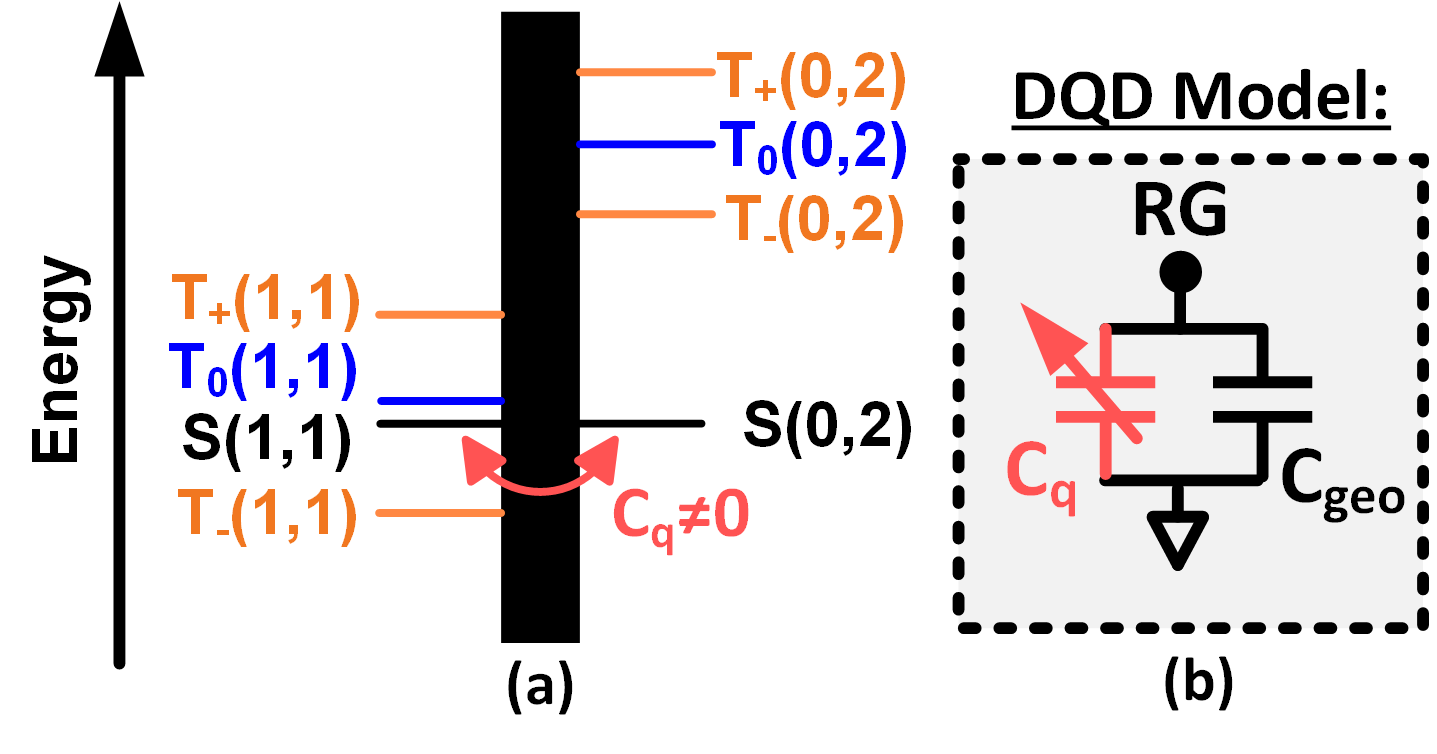} 
    \caption{(a) Sketch of the singlet and triplet energy levels of the DQD for the (1,1) and (0,2) charge configurations. Due to the smaller E\textsubscript{ST} of the (1,1) state compared to the (0,2) state, the lowest energy level of the DQD is T\textsubscript{-}(1,1) when $\epsilon=0$ \cite{hansonSpinsFewelectronQuantum2007}; (b) Equivalent lump model of the DQD under RF excitation. \label{fig:Singlet_Triplet_Model}}
\end{figure}

The energy levels of the singlet and triplet configurations for the (1,1) and (0,2) charge configurations are plotted together in Fig.\,\ref{fig:Singlet_Triplet_Model}(a). The sketch's left and right columns indicate the singlet and triplet energy levels for the (1,1) and (0,2) charge configurations, respectively. The figure illustrates when the DQD is at the interdot regime, equivalently described as the zero detuning condition ($\epsilon$\,=\,0). Here, the detuning parameter ($\epsilon$) is defined as the energy difference between the S(1,1) state and the S(0,2) state. In case the DQD collapses to a singlet state, an electron from one QD can freely tunnel to another at the zero detuning as the S(1,1) and the S(0,2) states have similar energy levels. In contrast, an electron cannot oscillate back and forth when the DQD is in a triplet state, as there are no overlapping triplet states when $\epsilon$\,=\,0.

As the electron moves between the QD sites due to an RF readout signal applied at RG, quantum capacitance is generated \cite{mizutaQuantumTunnelingCapacitance2017}. Consequently, the DQD can be modeled as two parallel capacitors during readout, as illustrated in Fig.\,\ref{fig:Singlet_Triplet_Model}(b), where the capacitance contribution comes from namely the geometric capacitance ($C_{geo}$) and the state-dependent quantum capacitance ($C_{q}$). Thus, modeling the readout behavior of the system relies on quantifying the impact of quantum capacitance, which is the main focus of the subsequent section.

\subsection{Quantum Capacitance Theory}
 As defined in \cite{Petersson_2010}, the quantum capacitance exhibited by the DQD is expressed as
\begin{equation}
\begin{aligned}
   C_{q} &= -\beta^2\frac{\partial^2 E_{\ket{\psi}}}{\partial \epsilon^2}, \label{eq:Cq}
\end{aligned}
\end{equation}
where $E_{\ket{\psi}}$ is the energy level of the singlet or triplet state of a DQD and $\beta$ is the lever arm of the DQD device, and is given by 
\begin{equation}
\begin{aligned}
   \beta &= |e| C_g / C_{\Sigma}, \label{eq:beta}
\end{aligned}
\end{equation}
where $|e|$ is the electron charge, $C_g$ is the capacitance of the readout gate, and $C_{\Sigma}$ is the total geometric capacitance of the DQD.

To calculate the quantum capacitance contribution from the singlet or triplet state expressed in (\ref{eq:Cq}), the energy level of the DQD for different detunings must be calculated in each respective configuration. This can be achieved by solving the DQD's Hamiltonian using the Fermi-Hubbard model\cite{hensgensQuantumSimulationFermi2017}. However, solving the total Fermi-Hubbard Hamiltonian of the DQD can be too complex and cumbersome. As we are only interested in the relevant features of the DQD under readout conditions, we simplify the DQD model and approximate the behavior by examining the Hamiltonian of a five-level system consisting of the ground and excited states of the S(1,1)-S(0,2) state and the three (1,1) triplet states. In general, the Hamiltonian for the S(1,1)-S(0,2) singlet state can be expressed by
\begin{align}
    H_{DQD,\ket{S}} &= \frac{\epsilon}{2}\tau_z + t_c \tau_x \label{eq:H_DQD},  
\end{align}
where $t_c$ is the tunnel coupling, and $\tau_{z,x}$ are the Pauli matrices \cite{nielsen_quantum_2010}. By solving the Hamiltonian with the time-independent Schrodinger equation, one can obtain the energy levels of the DQD for the singlet state as a function of detuning, expressed as
\begin{align} 
    E_{\ket{S_{g,e}}} &= \mp\frac{1}{2}\sqrt{\epsilon^2 + 4 t_c^2}. \label{eq:E_DQD} 
\end{align}

Furthermore, the energy levels of the (1,1) triplet state as a function of detuning can be approximated by \cite{mizutaQuantumTunnelingCapacitance2017}
\begin{align}
    E_{\ket{T_0}} &= \frac{\epsilon}{2} \label{eq:E_T0}, \\
    E_{\ket{T_{\pm}}} &= \frac{\epsilon}{2} \pm g\mu_B B, \label{eq:E_Tpm}
\end{align}
where $g$ is the electron gyromagnetic ratio, $\mu_B$ is the Bohr magneton constant, and $B$ is the applied magnetic field. 


\begin{figure}
    \centering
    \includegraphics[width=0.45\textwidth]{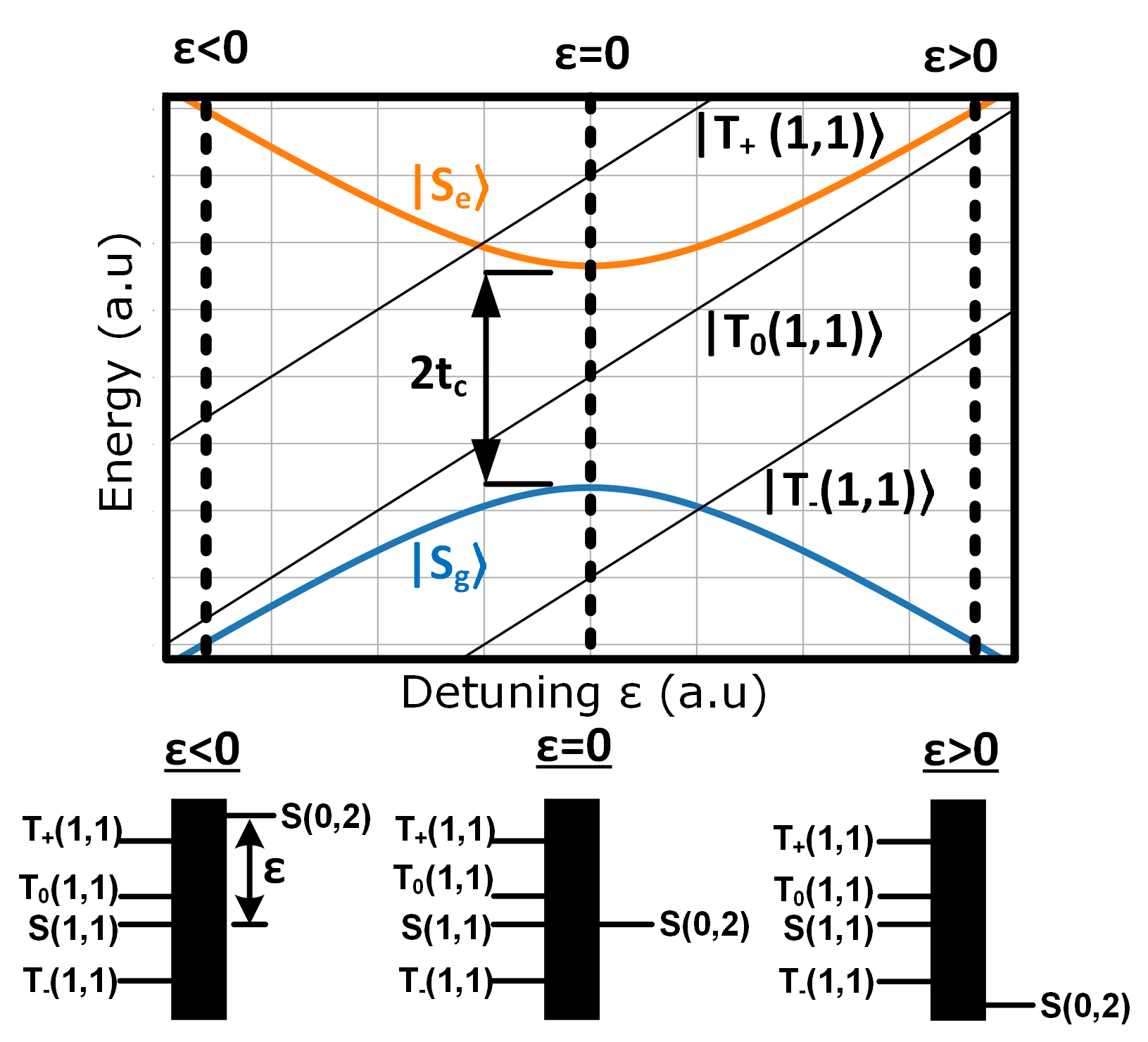} 
    \caption{Energy level diagram of a DQD based on the two-level quantum system approximation. Note that the triplet (0,2) energy levels are not plotted as they are significantly higher than the (1,1) triplet energy levels. \label{fig:Energy_level}}
\end{figure}

The energy levels of the singlet and triplet states expressed in (\ref{eq:E_DQD})-(\ref{eq:E_Tpm}) as a function of the detunings are plotted in Fig.\,\ref{fig:Energy_level}. By plugging the energy expression (\ref{eq:E_DQD})-(\ref{eq:E_Tpm}) to (\ref{eq:Cq}), the induced quantum capacitance of the singlet and triplet states can be respectively derived as
\begin{numcases}{C_q =}
 +\beta^2 \frac{2t_c^2}{(\epsilon^2+4t_c^2)^{3/2}} & \text{for $\ket{S_g}$,}\label{eq:Cq_singlet} \\
 -\beta^2 \frac{2t_c^2}{(\epsilon^2+4t_c^2)^{3/2}} & \text{for $\ket{S_e}$,} \label{eq:Cq_singlet_neg} \\
 0 & \text{for $\ket{T_0}$ and $\ket{T_\pm}$.} \label{eq:Cq_triplet}
\end{numcases}
Note that only the singlet state contributes an additional quantum capacitance during readout due to the curvature of the energy level. Moreover, the ground singlet state produces a positive quantum capacitance, whereas the excited singlet state produces a negative quantum capacitance. In contrast, no quantum capacitance contribution exists when the DQD is in the triplet states, as their energy levels are linearly dependent on $\epsilon$.

As previously mentioned, the right QD can be used as a reference spin for readout, in which its spin is initialized to a spin down and is used to compare the spin state of the left QD. Consequently, the two lowest energy levels used as the computational basis are the $\ket{S_g}$ and $\ket{T_-}$ states [see Fig.\,\ref{fig:Energy_Orbital_Diagram}]. For convenience, the two computational basis states used for the analysis are referred to as the $\ket{S}$ and the $\ket{T}$ state for the rest of the paper unless stated otherwise. 

Based on the developed equations and theory for quantum capacitance, we can discuss and model the DQD behavior during readout in the following subsection.

\subsection{Quantum Capacitance During Readout}
\begin{figure}
    \centering
    \includegraphics[width=0.43\textwidth]{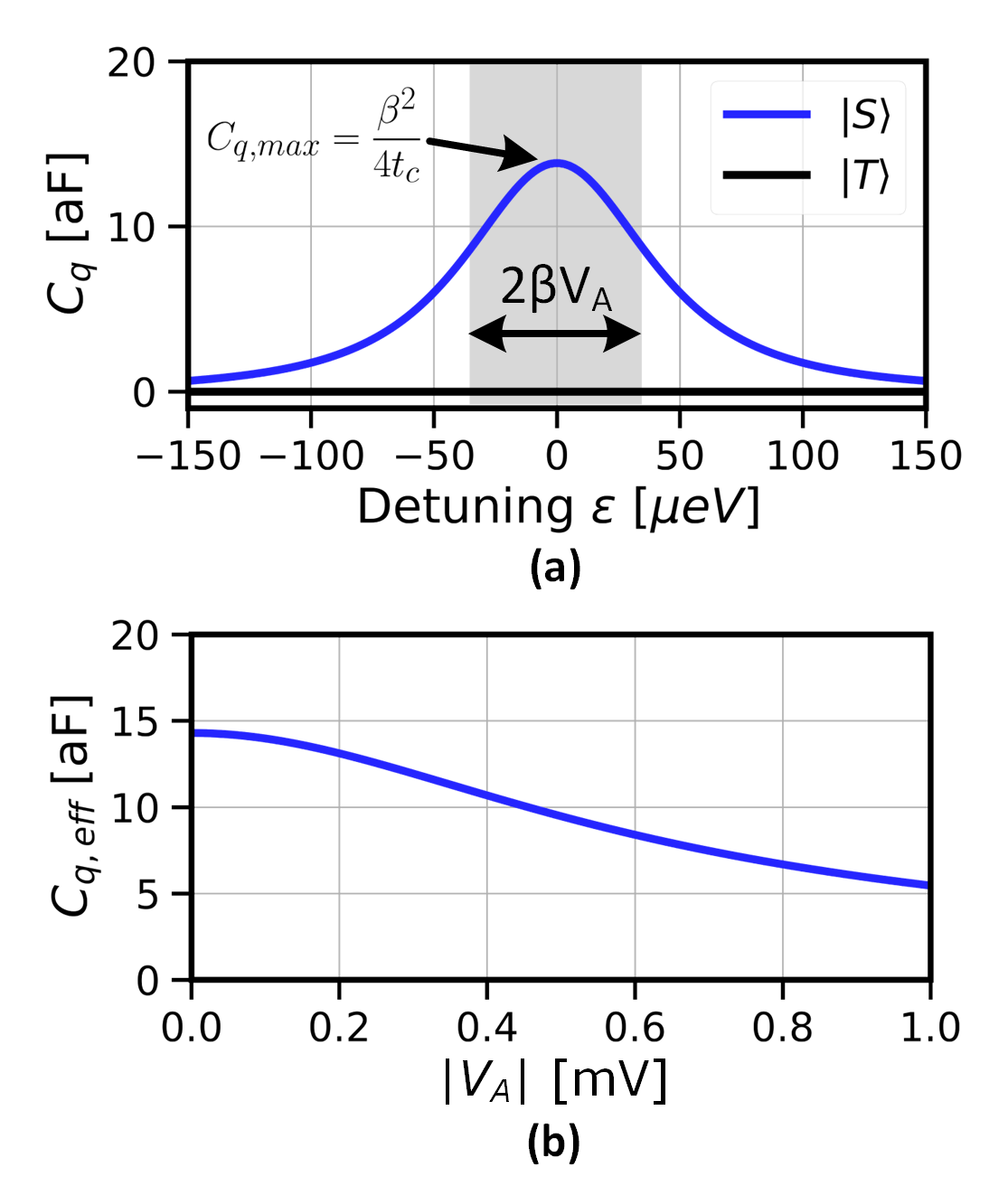} 
    \caption{(a) Instantaneous quantum capacitance of the DQD with respect to the detuning of the DQD; (b) Effective quantum capacitance versus the signal amplitude at the readout gate (RG). \label{fig:Cq_Plot}}
\end{figure}

In practice, the readout operation involves averaging the signal response of the system over many RF cycles to increase the SNR. Since the signal is averaged during readout acquisition, the quantum capacitance that the DQD contributes is also effectively averaged out. To illustrate this, based on (\ref{eq:Cq_singlet}), the quantum capacitance profile versus $\epsilon$ is illustrated in Fig.\,\ref{fig:Cq_Plot}(a). Under the assumption that the readout gate (i.e., RG) is driven by a signal in the form of $V_{A}\sin(\omega_{RF} t)$, the detuning value changes sinusoidally over each RF cycle:
\begin{align}
   \epsilon = \beta V_{A}\sin(\omega_{RF} t) \label{eq:driving_sig}, 
\end{align}
in which, the role of the lever arm ($\beta$) is to convert voltage quantities to energy, with $V_A$ denoting the signal's amplitude, and $\omega_{RF}$ as the angular frequency of the applied signal. Consequently, as shown in Fig.\,\ref{fig:Cq_Plot}(a), the value of the quantum capacitance varies along the shaded region due to the sinusoidal excitation. Due to the readout acquisition's averaging nature, the quantum capacitance's effective value seen during readout is the weighted average of the quantum capacitance value across the shaded region. By substituting (\ref{eq:driving_sig}) into (\ref{eq:Cq_singlet}), the effective quantum capacitance can be expressed as
\begin{align}
   C_{q,eff} &= \frac{1}{T_{int}} \int_0^{T_{int}} C_{q,\ket{S_{g}}}(\epsilon) \text{ d}t \bigg\rvert_{\epsilon\,=\,\beta V_{A}\sin(\omega_{RF} t)}. \label{eq:Cq_eff}
\end{align}
Provided that the integration time is an integer multiple of the RF readout signal's period (i.e., $T_{int}$ = $nT_{RF}$), a closed-form expression for $C_{q,eff}$ can be obtained,
\begin{align}
  C_{q,eff} = \frac{\beta^2 t_c \times E\left(2\pi, \frac{-\beta^2V_{A}^2}{4t_c^2}\right)}{2\pi (\beta^2V_{A}^2 + 4t_c^2)}, \label{eq:Cq_eff_2}
\end{align}
where, $E(\theta,k)$ is an elliptical integral expressed as
\begin{align}
  E(\theta,k) = \int^\theta_0 \sqrt{1-k^2sin\theta}d\theta.
\end{align}


The effective capacitance for different readout amplitudes is plotted in Fig.\,\ref{fig:Cq_Plot}(b), which is based on the result of (\ref{eq:Cq_eff_2}). A smaller readout excitation is observed to induce a larger effective quantum capacitance due to the smaller averaging window, which only captures the peak of the quantum capacitance profile. At small readout voltages, $C_{q,eff}$ converges to $\beta^2/4t_c$, in line with (\ref{eq:Cq_singlet}) when $\epsilon$\,=\,0. Also, note that a small readout excitation degrades the readout signal [see eq.\,(\ref{eq:SNR_theoretical_intro})], which may reduce the SNR for a given fixed $T_{int}$. Hence, a longer $T_{int}$ may be required to regain the same SNR. In contrast, a larger readout excitation implies a smaller effective quantum capacitance, which may, in effect, decrease the frequency shift of the response between the $\ket{S}$ and $\ket{T}$ states. This behavior is reflected in the state-separation factor between the $\ket{S}$ and $\ket{T}$ and consequently may lower the SNR as described in (\ref{eq:SNR_theoretical_intro}).

\subsection{Optimum Tunnel Coupling}
It should be noted that the tunnel coupling can significantly influence the behavior of the qubit during readout. When the tunnel coupling is set too low, the energy gap between the $\ket{S_e}$ band and the $\ket{S_g}$ band diminishes [see Fig.\,\ref{fig:Energy_level}]. As the $\ket{S_e}$ band gets closer to the $\ket{S_g}$ state, the system can transition between the $\ket{S_e}$ and $\ket{S_g}$ energy bands instead of staying in the $\ket{S_g}$ band, which is necessary for the system to realize quantum capacitance. For the assumption presented here to be valid, the system must be biased in the adiabatic limit, bounded by $(2t_c/h)/f_r>1$ \cite{mizutaQuantumTunnelingCapacitance2017}, where $f_r$ is the readout frequency and $h$ is the Planck constant. The system is more adiabatic and less likely to transition between the $\ket{S_e}$ and $\ket{S_g}$ state as the ratio between the tunnel coupling and the readout frequency $f_r$ increases. 

\begin{figure}
    \centering
    \includegraphics[width=0.43\textwidth]{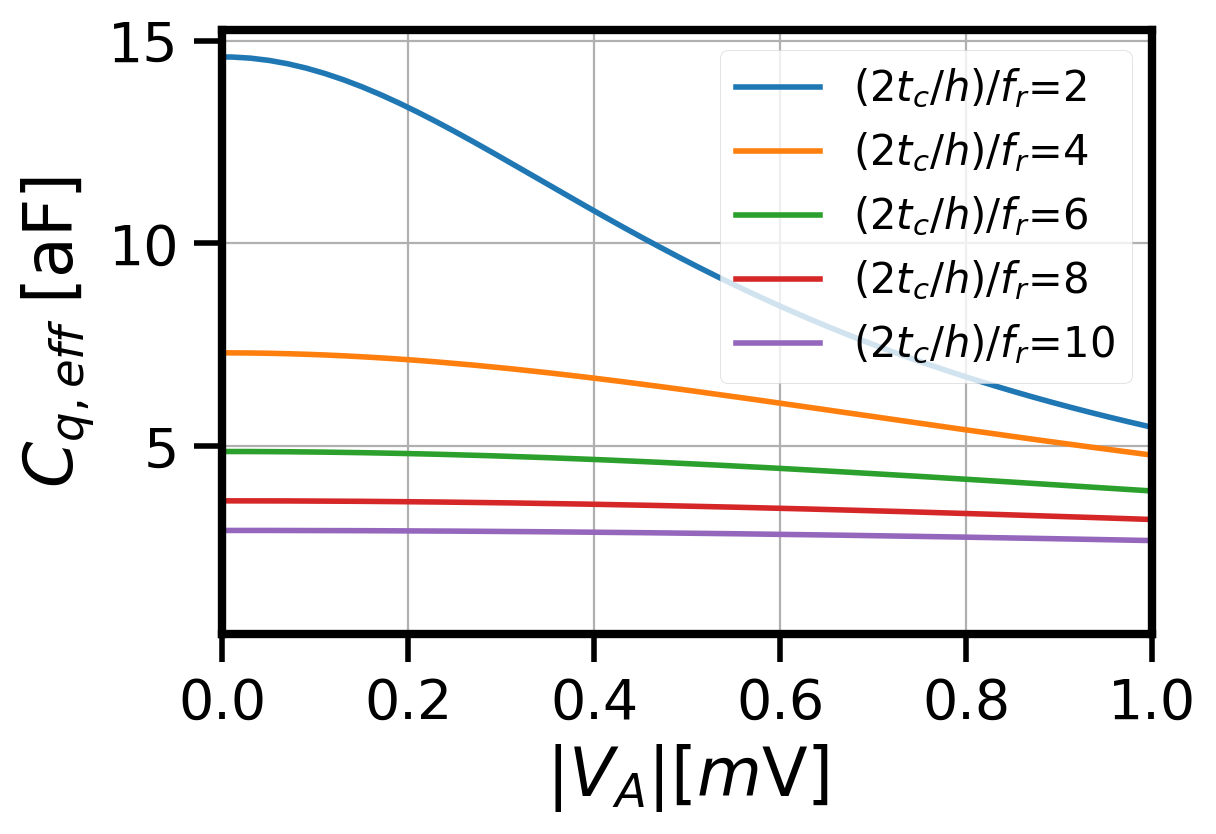} 
    \caption{$C_{q,eff}$ for different adiabaticity factors (i.e., $(2tc/h)/f_r$). 
    \label{fig:cq_vs_tc}}
\end{figure}

The $C_{q,eff}$ for different adiabaticity factors are plotted in Fig.\,\ref{fig:cq_vs_tc}. While it is desired to operate below the adiabatic limit, it is observed that larger quantum capacitance is observed by biasing the DQD closer to the limit, which can increase the sensitivity of readout. Based on the observation, the qubit must be biased close to the adiabatic limit (i.e., $(2t_c/h)/f_r = 2$) to maximize the quantum capacitance contribution.


\subsection{Lever Arm} 
The quantum capacitance can also be increased by means of the lever arm ($\beta$) to increase the readout sensitivity, as described by (\ref{eq:Cq_singlet}). The lever arm can be increased by designing the QD devices with quantum wells closer to the gate electrodes or by implementing the device in a silicon substrate, similar to a CMOS process \cite{Ansaloni_LETI_qubits, Tanttu_Silicon_qubits_2019}. While the approach aligns well with the future goal of monolithic integration with CMOS electronics, these devices tend to have faster decoherence times as the qubit is located closer to the gate dielectric interface \cite{Tanttu_Silicon_qubits_2019}. Thus, increasing the lever arm will need careful consideration as it influences multiple factors other than the readout performance.

\section{Readout Simulation} \label{sec:readout_sim}
\begin{figure}
    \centering
    \includegraphics[width=0.42\textwidth]{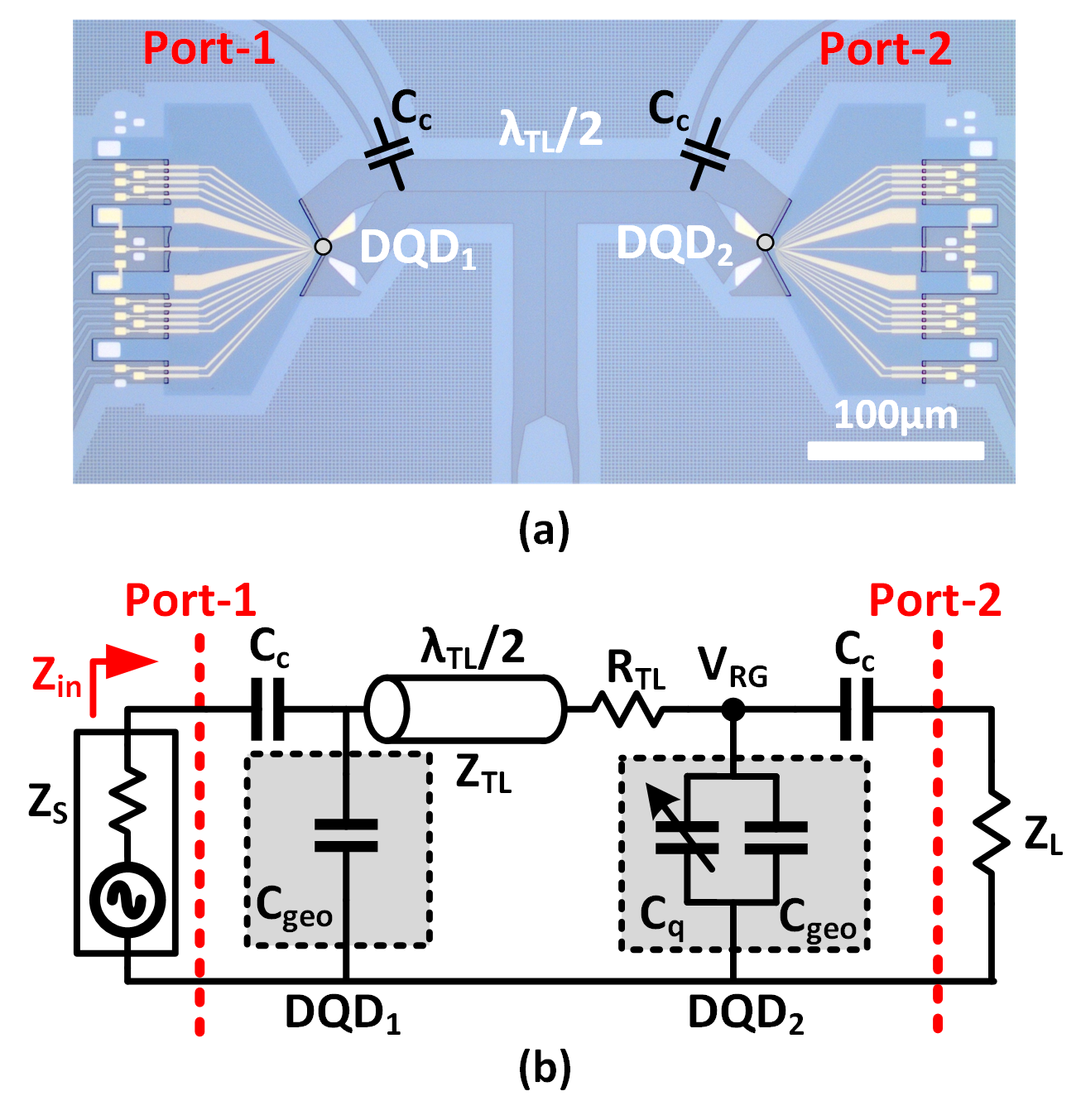} 
    \caption{(a) Micrograph of the qubit sample; (b) Equivalent circuit model of the sample. \label{fig:Sample_Model}}
\end{figure}

The quantum capacitance model introduced in the previous section can now be applied to a qubit sample that has been measured previously \cite{harvey-collard_circuit_2021}. The qubit micrograph is shown in Fig.\,\ref{fig:Sample_Model}(a). The sample, realized in \textsuperscript{28}Si/SiGe heterostructure, consists of 2 DQD sites separated by a 250$\,\mu$m half-wavelength superconducting transmission line fabricated with NbTiN material. The half-wavelength transmission line is AC coupled to the input and output ports (port-1 and port-2) through the coupling capacitors $C_{c}$. The DQD\textsubscript{2} is biased at the zero detuning regime, thus exhibiting a geometric and quantum capacitance during readout when DQD\textsubscript{2} is at the $\ket{S}$ state. Moreover, the DQD\textsubscript{1} is tuned in the (0,0) charge regime and thus only contains a fixed geometric capacitance. The sample is probed with an RF signal from port-1. The RF tone's corresponding phase or amplitude change can be detected by measuring its in-phase (I) and quadrature (Q) signal components at port-2 using an I/Q RX.

The complete electrical model of the sample is shown in Fig.\,\ref{fig:Sample_Model}(b). To model the state-dependent capacitance of DQD\textsubscript{2}, a voltage-dependent capacitor $C_{q}(V_{RG})$ is included in the circuit, and its value depends on the voltage at the readout gate [see Fig.\,\ref{fig:DQD_model}]. A resistor (i.e., $R_{TL}$) is also included in series to the transmission line in the schematic to model the system's loss, which can originate from resistive and dielectric losses in the sample \cite{Patrick_gate_filter_2020}. When comparing the simulation with the measurement results, $R_{TL}$ is used as a fitting parameter based on the measured resonator's bandwidth. In the simulation, the quantum capacitance contribution is realized by a lookup table (LUT) VerilogA model, relating different readout gate voltage levels to distinct $C_{q}$ values using (\ref{eq:Cq_singlet}) and DQD properties summarized in Tab. \ref{Tab:DQD_Property}.


\begin{table}[!ht]
\centering
\caption{System parameters for simulation and verification.}
\vspace{-0.1in}
\begin{tabular}[t]{lcc}
\hline
Quantity, Symbol & Value\\
\hline\hline
Tunnel coupling, 2$t_c$/h & 14.1 GHz \\
Lever arm, $\beta$ & 102 meV/V\\
Characteristic resonator impedance, $Z_{TL}$ & 4.5 k$\Omega$\\
Characteristic impedance of system,  $Z_{0}$ & 50 $\Omega$\\
Geometric DQD capacitance, $C_{geo}$ & 1.9 fF \\
Coupling capacitance, $C_{c}$ & 0.32 fF \\ 
Series resistance, $R_{TL}$ & 17 $\Omega$ \\
\hline
\end{tabular}
\label{Tab:DQD_Property}
\end{table}

\subsection{Frequency Shift} 

\begin{figure}
    \centering
    \includegraphics[width=0.37\textwidth]{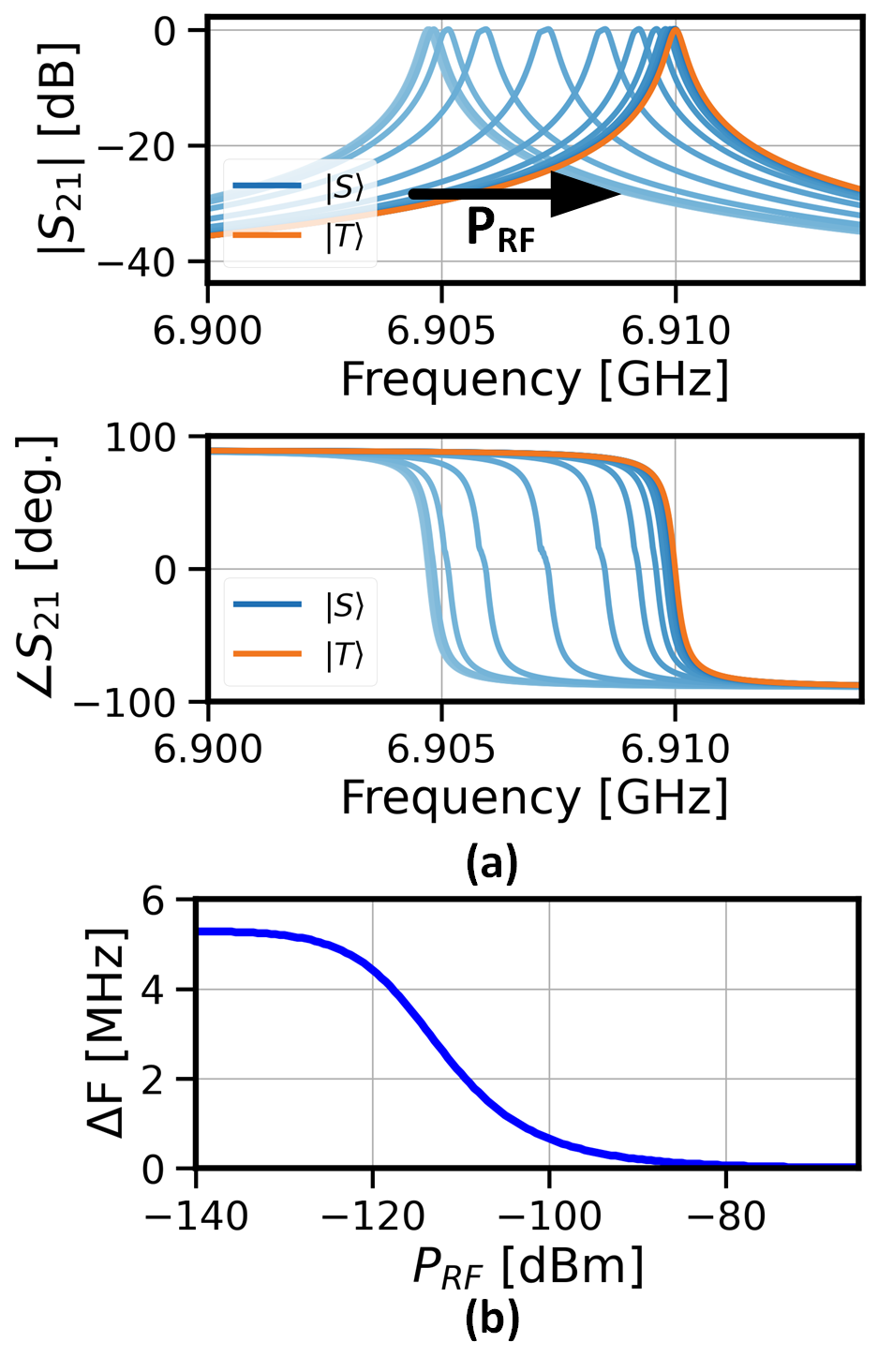} 
    \caption{(a) Simulated magnitude and phase response of the transmission gain (S\textsubscript{21}) of the readout system in Fig.\,\ref{fig:Sample_Model} for $P_{RF}$ of -140\,dBm to -60\,dBm in steps of 5\, dBm; (b) The frequency shift between the $\ket{S}$- and $\ket{T}$-state resonant frequencies for different RF input power. \label{fig:Ideal_res_sim}}
\end{figure}

The transmission (S\textsubscript{21}) behavior between port-1 and port-2 for different $P_{RF}$ values is plotted in Fig.\,\ref{fig:Ideal_res_sim}(a). The orange and blue lines in the plot indicate the transmission response when the DQD is in the $\ket{T}$ and the $\ket{S}$ state, respectively.


To gain more insight into the (S\textsubscript{21}) behavior, it is instructive to see the behavior of the input impedance of the sample. For simplicity, a large coupling capacitor $C_{c}$ and a lossless system  are first considered such that the input impedance of the sample can be approximated as
\begin{align}
   Z_{in} \approx Z_{TL,eff} \frac{Z_L + j  Z_{TL,eff}\tan{(\gamma l)}}{ Z_{TL,eff} + j Z_{L}\tan{(\gamma  l)}}, \label{eq:Zin_TL} 
\end{align}
where $l$ is the physical length of the transmission line, and $Z_{L}$ is the load impedance (equal to 50\,$\Omega$). $Z_{TL,eff}$ is the effective characteristic impedance of the transmission line defined as $\sqrt{{L_{TL}'}/{C_{TL,eff}'}}$, where $L_{TL}'$ is the inductance per unit length and $C_{TL,eff}'$ is the effective capacitance per unit length of the transmission line, including the capacitance of the DQD. Additionally, $\gamma$ is the phase constant defined as
\begin{align}
   \gamma = \omega \sqrt{L_{TL}'C_{TL,eff}'}, \label{eq:phase_constant}
\end{align}
where $\omega$ is the angular frequency where the expression is evaluated. On resonance, (\ref{eq:Zin_TL}) simplifies to 50\,$\Omega$ as $\tan(\gamma l)$ approaches 0. In effect, all the power provided by the source ($P_{RF}$) is delivered to port-1 and consequently also delivered to the load at port-2, resulting in a 0 dB response in the $|S\textsubscript{21}|$ plot [see Fig.\,\ref{fig:Ideal_res_sim}(a)]. Away from resonance, $\tan(\gamma l)$ tends to be large, thus $Z_{in}$ approaches $\sim$$Z_{TL,eff}^2/Z_L$. Due to the system's large $Z_{TL,eff}$, reflection at the input plane occurs, leading to an incomplete power transfer to $Z_L$. This reflection contributes to the observed bandpass response in the $|S\textsubscript{21}|$ plot.

The resonant condition requires that the $\gamma l$ term in (\ref{eq:Zin_TL}) be equivalent to $\pi$, such that $Z_{in}$ simplifies to 50\,$\Omega$. Consequently, the resonant frequency can be derived based on (\ref{eq:phase_constant}) and can be recast as 
\begin{align}
    \omega_{res} = \frac{\pi}{\sqrt{\left( L_{TL}' l \right) \times \left(C_{TL,eff}' l \right)}} \rightarrow \omega_{res} = \frac{\pi}{\sqrt{L_{TL}C_{TL,eff}}},
    \label{eq:res_freq}
\end{align}
where $L_{TL}$ and $C_{TL,eff}$ are the transmission line's total equivalent \textit{lumped} inductance and capacitance, respectively. Note that $C_{TL,eff}$, has two different loading conditions depending on the state of the DQD\textsubscript{2}, which can be described as
\begin{equation}
C_{TL,eff} = 
  \begin{cases}
  C_{TL,\ket{T}} = C_{TL} + 2C_{geo} & \text{$\ket{T}$ state,}\\
  C_{TL,\ket{S}} = C_{TL} + 2C_{geo} + C_{q} & \text{$\ket{S}$ state,}
  \end{cases}
  \label{eq:C_tl_eff}
\end{equation}
where $C_{TL}$ is the equivalent lumped capacitance of only the bare transmission line, excluding the DQD's contribution. 

When DQD\textsubscript{2} is in the $\ket{T}$ state, $C_{TL,eff}$ is only affected by $C_{geo}$. On the other hand, when DQD\textsubscript{2} is at the $\ket{S}$ state, $C_{TL,eff}$ is affected by $C_{geo}$ and $C_{q}$, which leads to a lower $\omega_{res}$, as observed from Fig.\,\ref{fig:Ideal_res_sim}(a). Additionally, when the system is subjected to a strong RF readout signal, $C_{q}$ is averaged over a larger span. This results in a lower $C_{q,eff}$, causing a smaller resonant frequency separation between the $\ket{S}$ and the $\ket{T}$ response.

To further increase the separation of the resonant frequency between the $\ket{T}$ and $\ket{S}$ state, it is necessary to increase the $C_{TL,\ket{S}}/C_{TL,\ket{T}}$ capacitance ratio. Hence, $C_q$ must be maximized while keeping $C_{TL}$ and $C_{geo}$ low. A higher $C_q$ can be achieved by increasing the lever arm of the device with the drawbacks mentioned in Section II. On the other hand, $C_{TL}$ can be kept low by designing the transmission line with a large characteristic impedance.

The simulated resonant frequency shift difference (i.e., $\Delta F$) of the $\ket{S}$-state relative to the $\ket{T}$-state for different input probe power is summarized in Fig.\,\ref{fig:Ideal_res_sim}(b). At low readout power, a $\sim$5\,MHz frequency shift is seen. However, the frequency shift becomes smaller at higher probe power as $C_{TL,\ket{S}}$ approaches $C_{TL,\ket{T}}$, which aligns with the presented theory.

\subsection{Readout Signal and Probe Frequency} 


\begin{figure}
    \centering
    \includegraphics[width=0.48\textwidth]{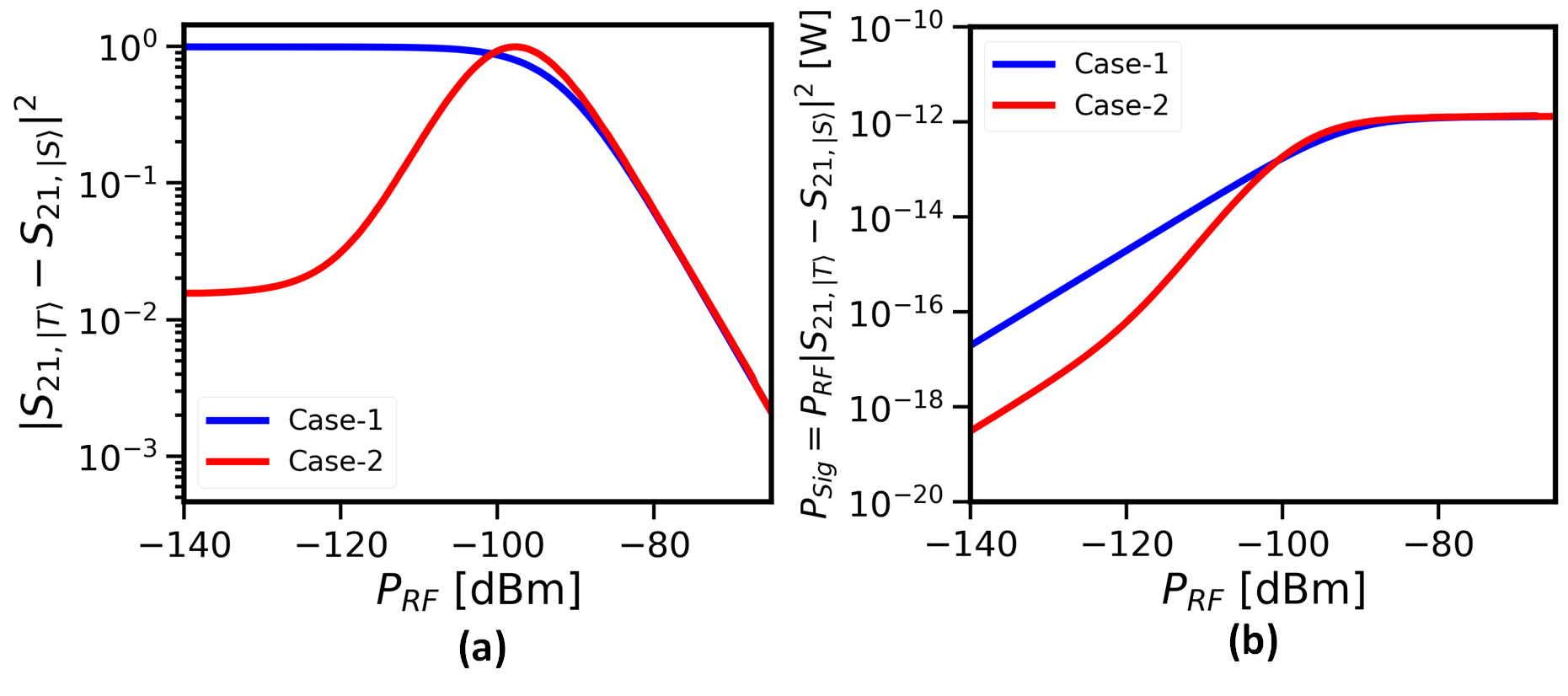} 
    \caption {Simulated readout response: (a) the state-separation factor; (b) the received power at port-2 of the transmission line (i.e., $P_{Sig}$). \label{fig:fr_resp_comparison}}  
\end{figure}

To determine the readout SNR, the readout signal's power at port-2 must be calculated. However, the behavior of the readout signal depends on the frequency at which the system is probed. This section explores the behavior of the readout signal in two scenarios: (1) when the system is probed at the resonant frequency of state $\ket{T}$ (i.e., 6.91\,GHz) leading to a BASK response, and (2) at the frequency halfway in between the resonant frequency response of the $\ket{S}$ and $\ket{T}$ state, leading to a BPSK response.

As expressed in (\ref{eq:P_sig_intro}), the readout signal's power ($P_{sig}$) is defined as the product of the state-separation factor and the power applied to the readout sample. The simulated state-separation factor (i.e., $|{S_{21,\ket{T}}}(\omega_r)$\,-\,${S_{21,\ket{S}}}(\omega_r)|^2$) for both cases are plotted in Fig.\,\ref{fig:fr_resp_comparison}(a). In case-1, the state-separation factor is shown to be approximately 1 for low readout probe power. In this condition, the ${S_{21,\ket{S}}}$ term is significantly smaller than the ${S_{21,\ket{T}}}$ term when probed at the resonance of the $\ket{T}$ state. Consequently, the $\ket{S}$ response is attenuated during readout compared to the $\ket{T}$ response, resembling a BASK readout signal, as discussed previously. In case-2, the state-separation factor is shown to be smaller than case-1 at low readout power. In this condition, both the ${S_{21,\ket{T}}}$ and ${S_{21,\ket{S}}}$ terms are generally small when probed between the $\ket{T}$ and $\ket{S}$ resonant frequencies. This results in a readout response that is both attenuated and exhibits a 180-degree phase difference between the $\ket{T}$ and $\ket{S}$ readout response, akin to a BPSK signal. Interestingly, as the applied input power increases, the state-separation factor converges, reflecting that the resonant frequencies of the $\ket{T}$ and $\ket{S}$ states converge.

The readout signal's power is shown in Fig.\,\ref{fig:fr_resp_comparison}(b). Remarkably, the plot shows a saturation in the readout signal power due to the decrease in the state-separation factor with the application of higher readout probe power. As the readout signal power dictates the readout SNR [see eq.\,(\ref{eq:SNR_theoretical_intro})], the simulation implies that there is a maximum achievable SNR for a given qubit property. Interestingly, in a different way and based on the time-dependent charge population in the DQD under RF excitation, \cite{maman_charge_2020} also theoretically predicted the saturation in the readout signal. However, the work in \cite{maman_charge_2020} did not consider the power-dependent frequency shift and did not present any experimental results to support the theory.

From this observation, performing a readout shown by case-1 is generally recommended, as it yields a larger readout signal response even at lower probe power, leading to a higher SNR even at lower probe power. Furthermore, from a practical point of view, readout at the bare resonant peak of the resonator is simpler as the readout frequency is fixed and does not need to be changed to account for the power-dependent resonant shift. Based on this reasoning, the experiments in this paper follow the readout behavior shown by case-1.

\subsection{Noise Estimation}
The power spectral density of the system's noise ($N_0$) at the input of the readout chain can be expressed as $kT_{N} $(\,W/Hz), where $k$ is the Boltzmann constant, and $T_N$ is the effective noise temperature of the readout system. Considering an ambient temperature of $T_{amb}$ for the DQD sample, $T_N$ may be estimated by
\begin{align}
    T_{N} = T_{amb} + T_{TWPA} + \frac{T_{LNA}}{G_1} + \frac{T_{RX}}{G_1 G_2},
\end{align}
where $T_{TWPA}$, $T_{LNA}$, and $T_{RX}$ are the noise temperatures of the traveling wave parametric amplifier (TWPA), low noise amplifier (LNA), and the RX, respectively, while $G_1$ and $G_2$ are the gain of the TPWA and LNA, respectively. Considering a TWPA gain of 28\,dB ($G_1$), the readout system's noise performance is mainly dictated by the TWPA. Assuming that the TWPA is quantum-limited, its noise temperature can be approximated by $T_{TWPA}$\,=\,$hf_r/k$ \cite{kerr1997receiver}. For an $f_r$ of 6.91\,GHz, and a $T_{amb}$ of 20\,mK, the total noise temperature of the readout architecture ($T_N$) is estimated to be 350\,mK. 

\subsection{Normalized Signal-to-Noise Ratio}
\begin{figure}
    \centering
    \includegraphics[width=0.39\textwidth]{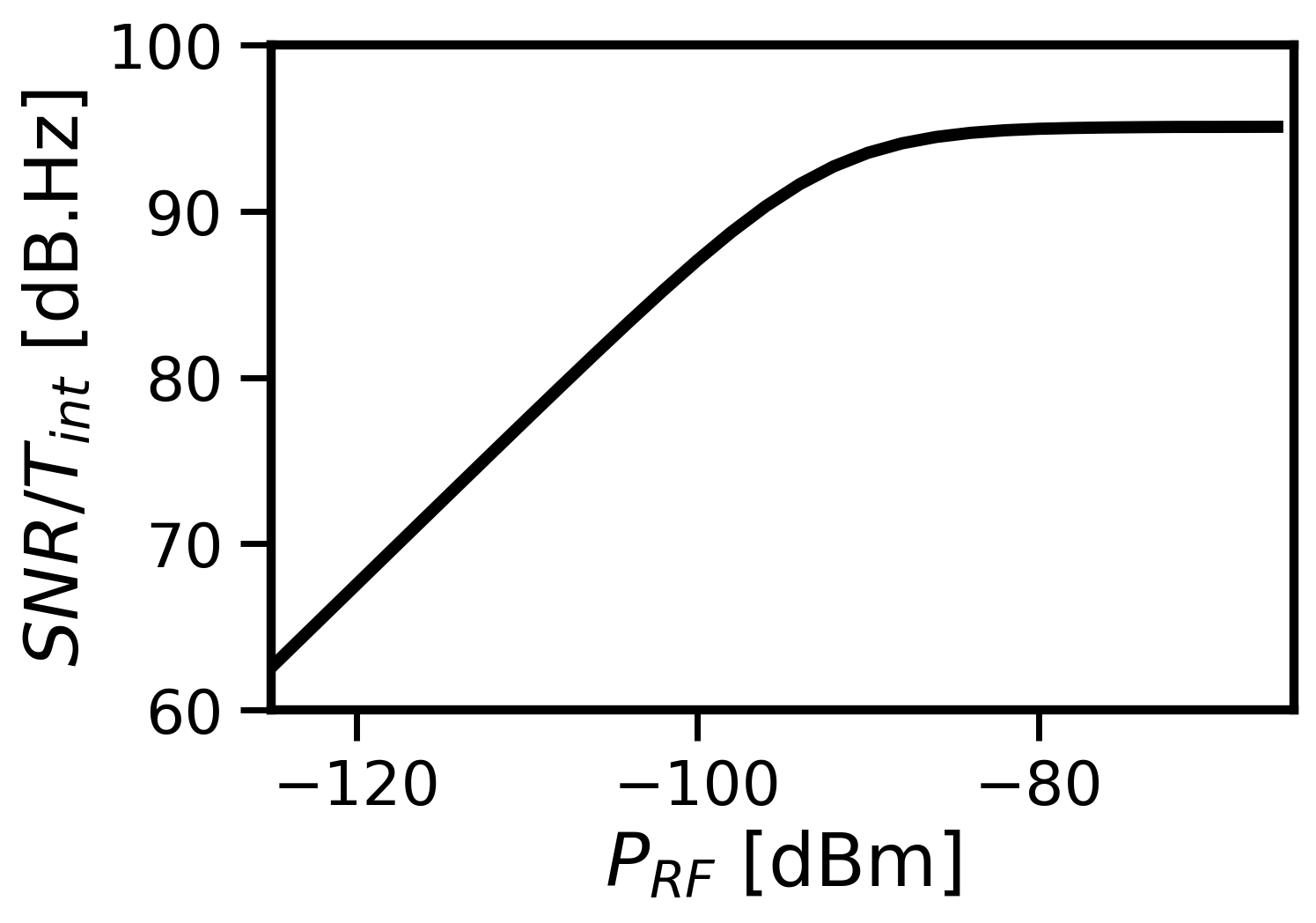} 
    \caption {Simulated normalized SNR, assuming a quantum-limited readout chain, and $T_{N}$\,=\,350\,mK. \label{fig:Ideal_SNR_N}} 
\end{figure}

The behavior of the readout SNR can now be discussed. However, to properly evaluate the trade-off between $P_{RF}$ and $N_0$, it is far more instructive to define the normalized SNR: SNR\textsubscript{N}\,=\,$P_{Sig}/N_0$\,=\,SNR/$T_{int}$. In this way, we can compare the performance of the readout scheme independent of the chosen $T_{int}$. The simulated SNR\textsubscript{N}, when assuming a $T_{N}$ of 350\,mK and considering all the system parameters in Tab.\,\ref{Tab:DQD_Property}, is plotted in Fig.\,\ref{fig:Ideal_SNR_N}. The SNR\textsubscript{N} firstly increases linearly with the probe power but finally saturates at a maximum of 95\,dB$\cdot$Hz. Considering a $T_{int}$ of 1\,$\mu$s, and assuming a quantum-limited readout chain, a theoretical maximum SNR of 35\,dB is achievable.

With the theoretical framework for gate-based RF readout established, the measurement verification of the proposed model can now be carried out.


\section{Measurements} \label{sec:Measurements}

\begin{figure}
    \centering
    \includegraphics[width=0.44\textwidth]{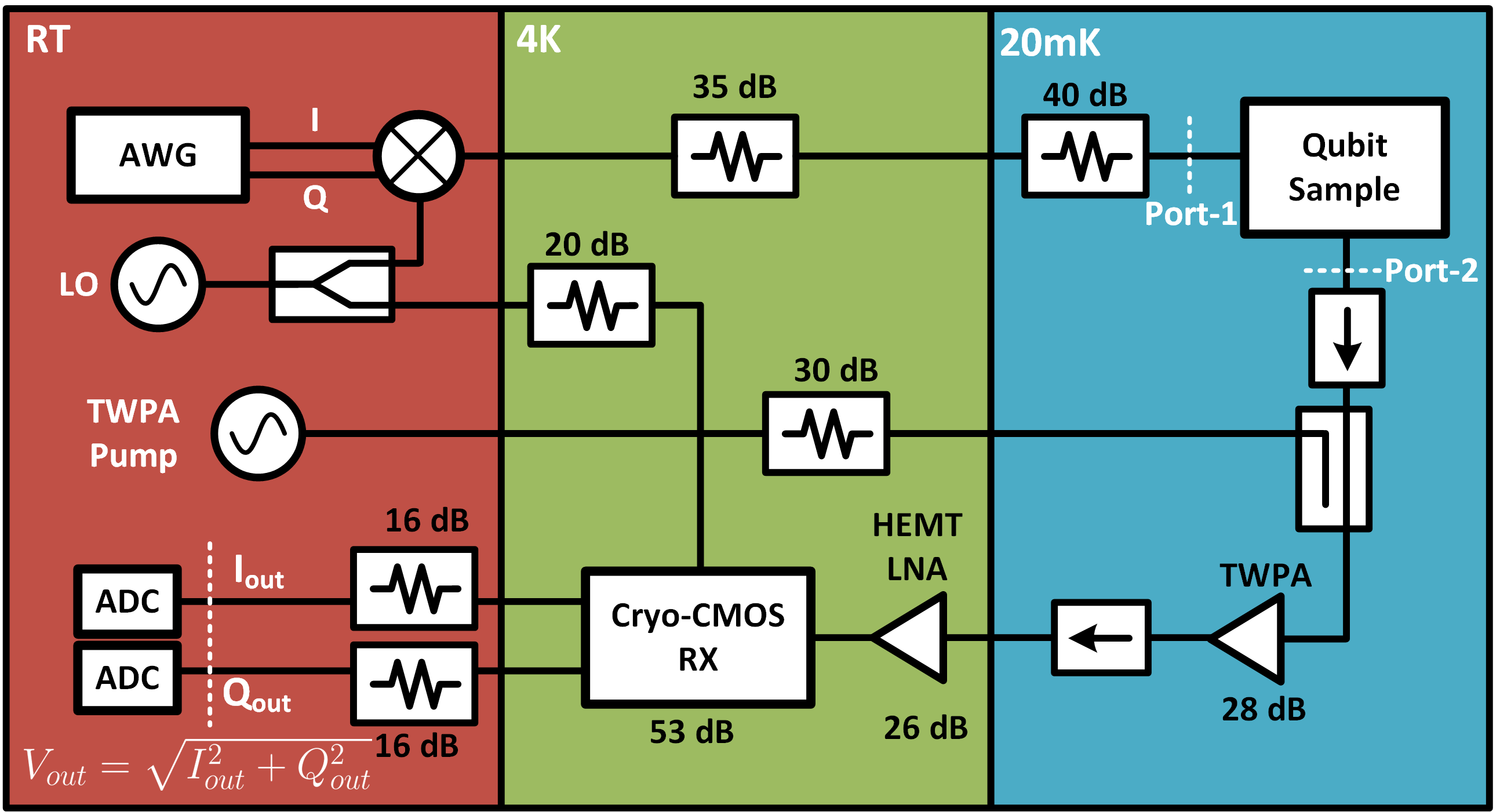} 
    \caption {Measurement setup used to verify the quantum capacitance behavior. \label{fig:Qubit_Setup_4K}}
\end{figure}

\subsection{Measurement Setup} 
The complete measurement setup is shown in Fig.\,\ref{fig:Qubit_Setup_4K}, similar to the ones used in \cite{prabowo136to8GHz17mW2021, harvey-collard_circuit_2021, Patrick_gate_filter_2020}. The sample is mounted at the mixing chamber plate and cooled down to an ambient temperature of 20\,mK. Alongside the qubit sample, a TWPA with 28\,dB gain is installed to minimize the noise of the readout chain, thus allowing one to measure a low-power readout signal at various $T_{int}$. The TWPA pump is generated at room temperature (RT) and fed to the TWPA by a coupler at the mixing chamber plate. An isolator is used between the qubit sample and the coupler to isolate the qubit from reflections due to the large TWPA pump tone. At the 4\,K plate, a high electron mobility transistor (HEMT) LNA with a 26\,dB of gain is used to amplify the readout signal further. An additional CMOS RX chip is mounted on this plate for signal downconversion and amplification \cite{prabowo136to8GHz17mW2021}. At RT, two off-the-shelf 1\,GS/s 8-bit ADCs are used to quantize both the I/Q signals, allowing for further digital signal processing at the baseband. A single LO source at RT is also used to drive the two mixers that upconvert and downconvert the readout signal to ensure phase synchronization. Moreover, a 75\,dB attenuation in the input RF line is intentionally employed to ensure a proper thermal noise level at the input of the sample at the mixing chamber plate. 

\subsection{Charge Stability Diagram}
\begin{figure}
    \centering
    \includegraphics[width=0.44\textwidth]{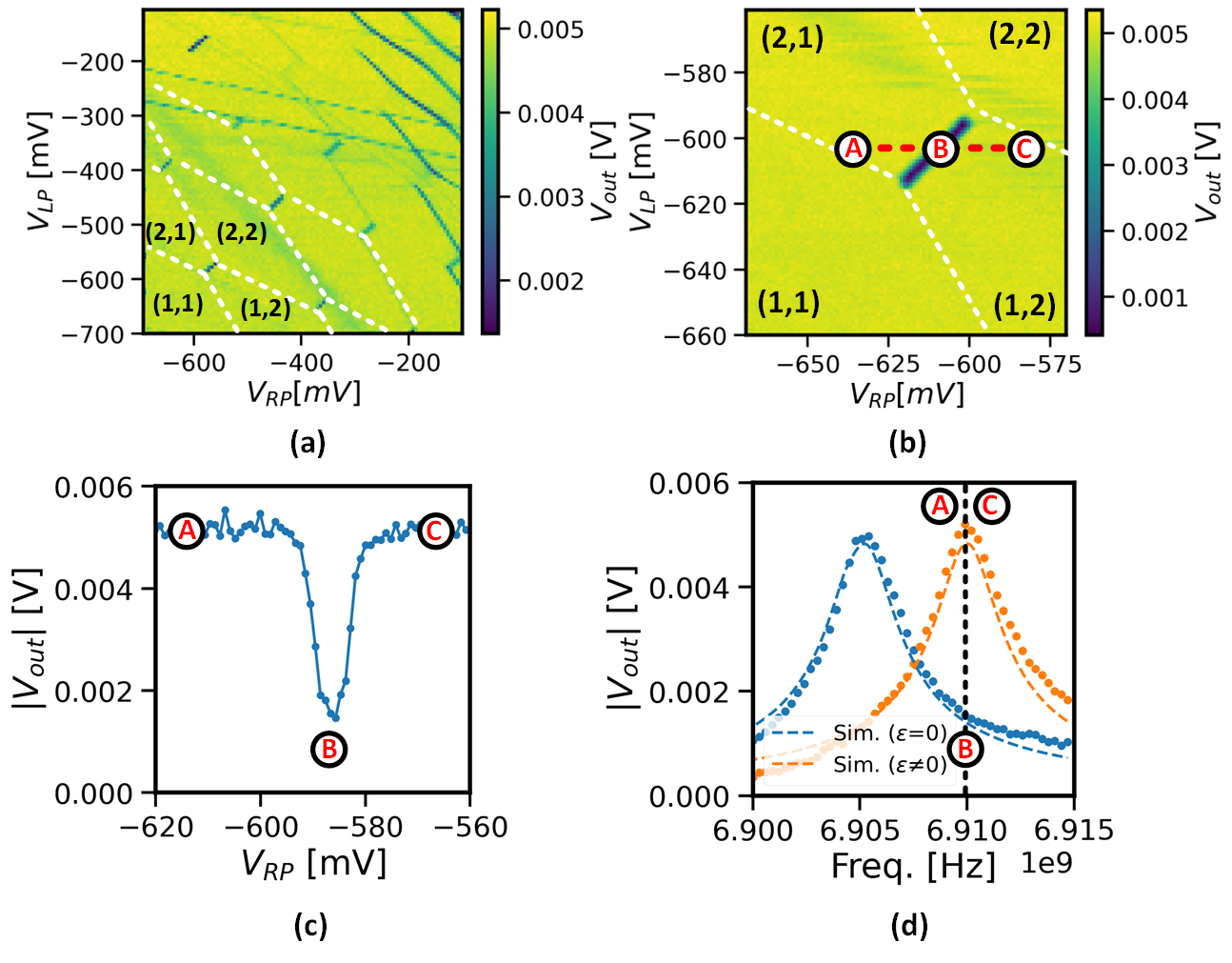} 
    \caption {(a) Measured charge stability diagram of DQD\textsubscript{2}. $|V_{out}|$ obtained from $I_{out}$ and $Q_{out}$ voltages at the RT plane [see Fig.\,\ref{fig:Qubit_Setup_4K}]; (b) Zoomed-in charge stability diagram at (2,1)-(1,2) charge regime. The next two sub-figures are measured across the red dashed line: (c) $|V_{out}|$ versus $V_{RP}$; (d) $|V_{out}|$ versus frequency at zero and non-zero detuning. \label{fig:Charge_stab}}
\end{figure}

The stability diagram of the sample is measured and shown in Fig.\,\ref{fig:Charge_stab}(a). The diagram is obtained when the qubit sample is read out at the bare resonance frequency (i.e., 6.91\,GHz) for different $V_{LP}$ and $V_{RP}$ values. A clear distinction between different charge states can be observed, especially at lower charge population regimes.


The behavior of $C_{q,eff}$ is verified in the (2,1)-(1,2) charge state. Considering that the quantum capacitance is generated by forcing an electron to oscillate across the DQD site under RF excitation, the behavior in the (2,1)-(1,2) regime is similar to the (1,1)-(0,2) regime discussed previously since the experiment isolates only the quantum capacitance contribution behavior. The charge stability scan around the (2,1)-(1,2) charge state is shown in Fig.\,\ref{fig:Charge_stab}(b). To understand the effect of the quantum capacitance during readout, a line cut is taken across the interdot crossing, and its corresponding result is depicted in Fig.\,\ref{fig:Charge_stab}(c). At point (A), the DQD is not loaded by the quantum capacitance as it is far from the zero detuning ($V_{RP}$\,=\,-620\,mV). As shown previously, averaging the quantum capacitance well away from the zero detuning results in a net zero effective quantum capacitance, which emulates the behavior of the $\ket{T}$ state. In contrast, point (B) is located where the system is at zero-detuning ($V_{RP}$\,=\,-590\,mV), where the maximum quantum capacitance contribution can be observed. Due to the quantum capacitance contribution, the resonator's resonance frequency shifts to a lower frequency, which, in effect, lowers the observed output voltage. As more positive $V_{RP}$ is applied to the DQD ($V_{RP}$\,=\,-560\,mV), the system returns to the response observed at (A).

The shift in the resonance frequency of the resonator can also be observed in Fig.\,\ref{fig:Charge_stab}(d). The data is obtained by monitoring the magnitude of the downconverted I/Q voltages (i.e., $|V_{out}|$) when the frequency of the probe signal is swept from 6.9\,GHz to 6.915\,GHz. For comparison, the simulation results are also included in Fig.\,\ref{fig:Charge_stab}(d) using the dotted lines, which show a good agreement with the measurements. The orange line in the figure depicts the response when the system is away from zero detuning ($V_{RP}$\,=\,-620\,mV), while the blue line depicts the system's response at zero detuning ($V_{RP}$\,=\,-590\,mV). The mapping of points (A), (B), and (C) from Fig.\,\ref{fig:Charge_stab}(c) are also labeled in Fig.\,\ref{fig:Charge_stab}(d) for clarity. As predicted from the simulation, a 5\,MHz frequency shift is observed for the zero-detuning response at very low power.

\subsection{Frequency Shift}

\begin{figure}
    \centering
    \includegraphics[width=0.4\textwidth]{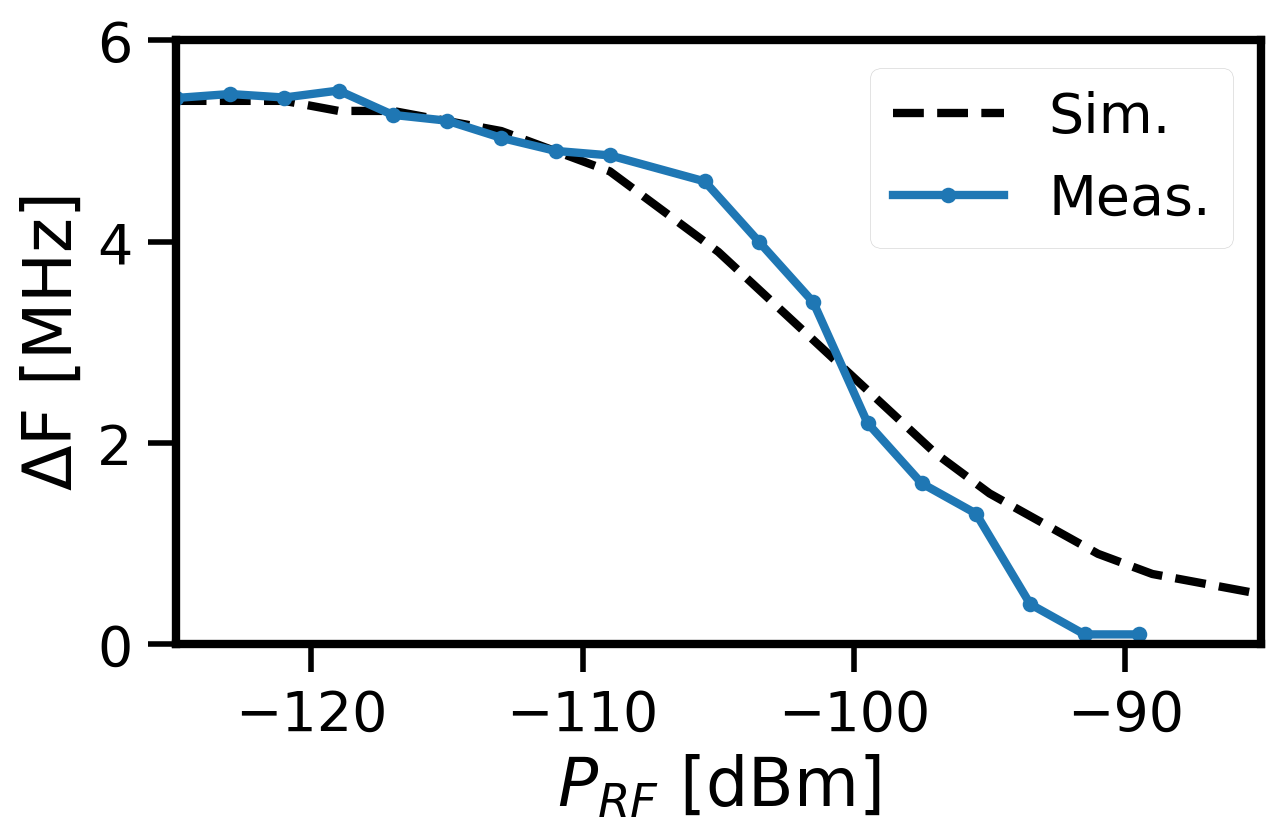}
    \caption {Observed frequency shift in simulation and measurements for different $P_{RF}$.\label{fig:freq_shift}}
\end{figure}

Given that the quantum capacitance cannot be measured directly, we can only observe the manifestation of the quantum capacitance through the frequency shift response of the sample. To verify this, the frequency at which the maximum transmission is observed is tracked at zero detuning and non-zero detuning conditions while sweeping the probe tone power from -125\,dBm to -89\,dBm, referred to the port-1 plane in Fig.\,\ref{fig:Qubit_Setup_4K}(c). The frequency difference ($\Delta F$) is defined as the difference between the resonance peaks of the non-zero and zero detuning responses. As shown in Fig.\,\ref{fig:freq_shift}, the frequency shift of the DQD decreases for higher input power, following the trajectory dictated by the developed theory and simulation results. Interestingly, the frequency shift reaches 0 faster than in the simulation. This behavior is discussed in the following section.

\subsection{Readout Signal}


\begin{figure}
    \centering
    \includegraphics[width=0.40\textwidth]{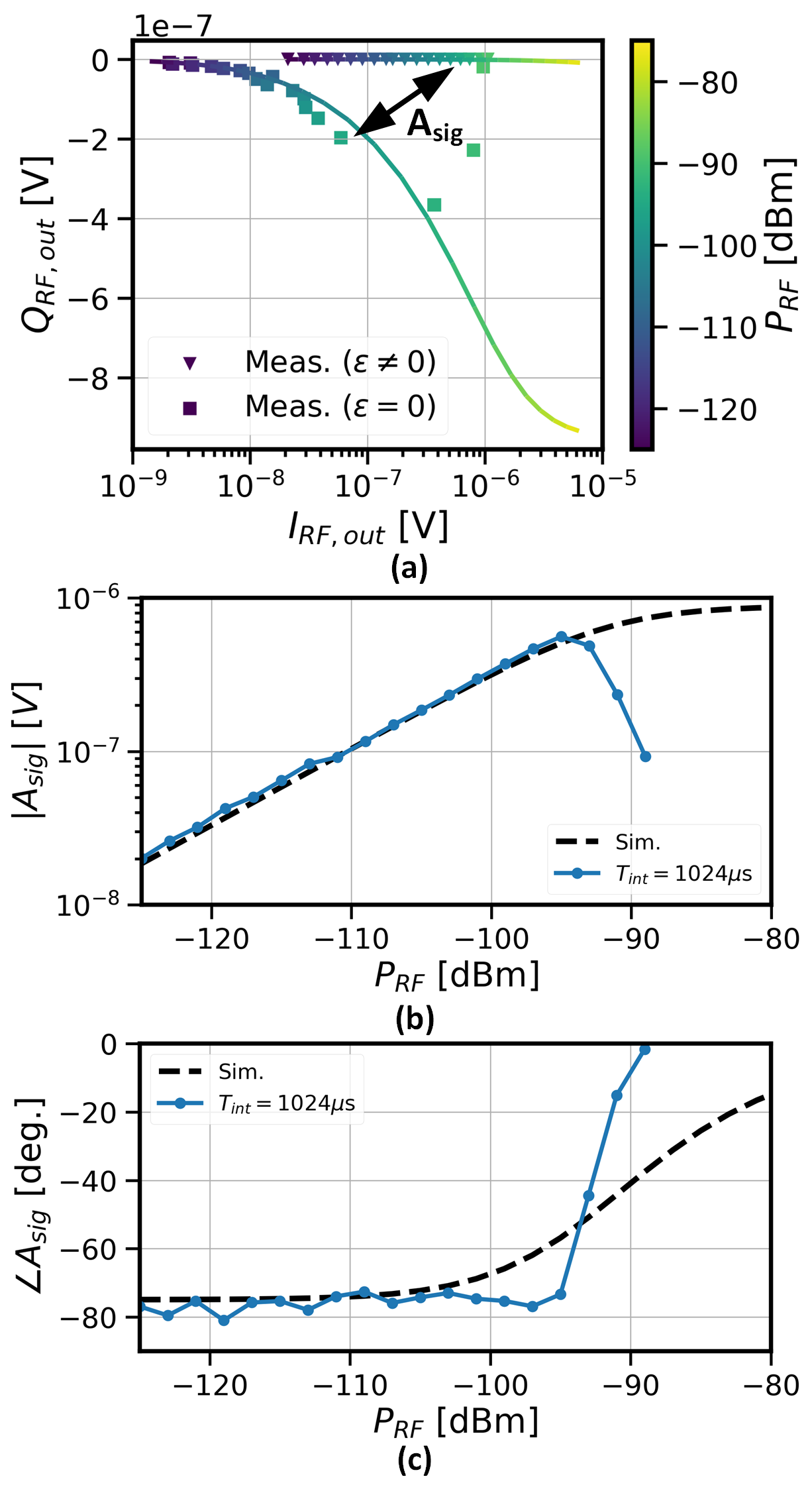} 
    \caption {(a) I/Q trajectory of readout signal for different $P_{RF}$ at zero and non-zero detuning regimes. The colored graded line indicates the expected I/Q trajectory from the simulation, while the triangle and the square data points indicate the measured value at non-zero and zero detuning, respectively. (b) The magnitude and (c) phase response of $A_{sig}$.\label{fig:A_sig_char}}
\end{figure}

The simulated and measured readout I/Q responses for zero and non-zero detunings are illustrated in Fig.\,\ref{fig:A_sig_char}(a). The data is obtained by referring the measured I/Q data sets to the port-2 plane when the system is probed at the readout frequency of $\omega_r/2\pi$\,=\,6.91 GHz. The triangle and square points indicate the non-zero and zero detuning responses, respectively. Note that the solid line indicates the simulated I/Q voltages, and the measured data points are color-graded based on the readout power applied to the sample. Based on Fig.\,\ref{fig:A_sig_char}(a), the I/Q points resemble an amplitude-modulated signal at lower readout power, where the zero-detuning response is close to the origin of the I/Q plane due to the attenuation of the shifted resonator response. At higher readout power, the zero-detuning I/Q response acquires an extra phase component as the resonant frequency of the zero-detuning response gets closer to the bare-resonance frequency of 6.91\,GHz. The distance between the two responses is also reduced for the higher power, implying a decreased readout signal amplitude at a larger probe power.

As explained in Section I and shown in Fig.\,\ref{fig:A_sig_char}(a), the readout signal ($A_{sig}$) is the distance between the qubit's responses in the constellation diagram. The magnitude and phase of $A_{sig}$ are respectively shown in Fig.\,\ref{fig:A_sig_char}(b) and Fig.\,\ref{fig:A_sig_char}(c). The simulated and measured $|A_{sig}|$ and $\angle A_{sig}$ are seen to be in good agreement up to $P_{RF}$ of $\sim$-100\,dBm. Beyond this point, a noticeable difference between simulation and measurement is observed, especially the measured $\angle A_{sig}$. Compared to the simulation, at large probe power, the measured $\angle A_{sig}$ approaches 0 degrees faster, and the measured $|A_{sig}|$ does not saturate but reduces significantly. Both magnitude and phase behaviors imply that the resonant frequency of the zero detuning response becomes the same as the bare resonance frequency of the resonator, consistent with the qualitative trend observed for $\Delta F$ in Fig.\,\ref{fig:freq_shift}. 
 
The discrepancy between the simulation and the measurement results at higher probe power levels can be attributed to several factors. The primary factor contributing to the reduced readout signal amplitude at large probe power levels is suspected to be associated with Landau-Zener transitions \cite{Zener_1932}. At a very large excitation, the rate at which the system's detuning evolves with time (i.e., $\partial \epsilon/\partial t$) is fast enough such that the evolution is no longer adiabatic. Hence, the system can not be examined with a time-independent Hamiltonian, as initially assumed. In this regime, the system can undergo transitions between the $\ket{S_g}$ and $\ket{S_e}$ energy bands, as illustrated in Fig.\,\ref{fig:Energy_level}, leading to a reduced readout response. Interestingly, this observation is consistent with the numerical simulation shown in \cite{maman_charge_2020}, which predicts that the quantum capacitance contribution can be reduced to 0 when the system is assumed to be non-adiabatic. In addition to the Landau-Zener transitions, the reduction in the readout amplitude may also be caused by the resonator's self-heating, which can cause the resonator's quality factor to degrade at a very large excitation. 

\subsection{Readout Noise}
The standard deviation of $|V_{out}|$ can be used to estimate $T_N$ experimentally. By measuring the standard deviation of $|V_{out}|$ at the non-zero detuning regime between a $V_{RP}$ of -570\,mV to -560\,mV (i.e., Point C in Fig.\,\ref{fig:Charge_stab}(c)), the integrated noise voltage can be calculated. Mathematically, this is equivalent to $\sqrt{kT_N\cdot 50/T_{int}}$\,(V). Experimentally, $T_N$ is estimated to be 460\,mK when referring to the input of the readout chain. 

The measured $T_{N}$ is close to what was previously observed for a readout chain incorporating TWPA amplifiers \cite{Macklin_2015, Simbierowicz_2021, white_traveling_2015}. However, it is 1.3$\times$ higher than the previous theoretical estimation (i.e., 350\,mK). The discrepancy between the theoretical and measured values may stem from different contributions. First, $T_{N}$ is estimated by always assuming that $T_{amb}$ is 20\,mK, which may be higher in reality as reflected by the electron temperature. Moreover, the estimation did not account for the losses of the isolator and coupler in front of the TWPA. 

\subsection{Normalized Signal-to-Noise Ratio}
 

\begin{figure}
    \centering
    \includegraphics[width=0.47\textwidth]{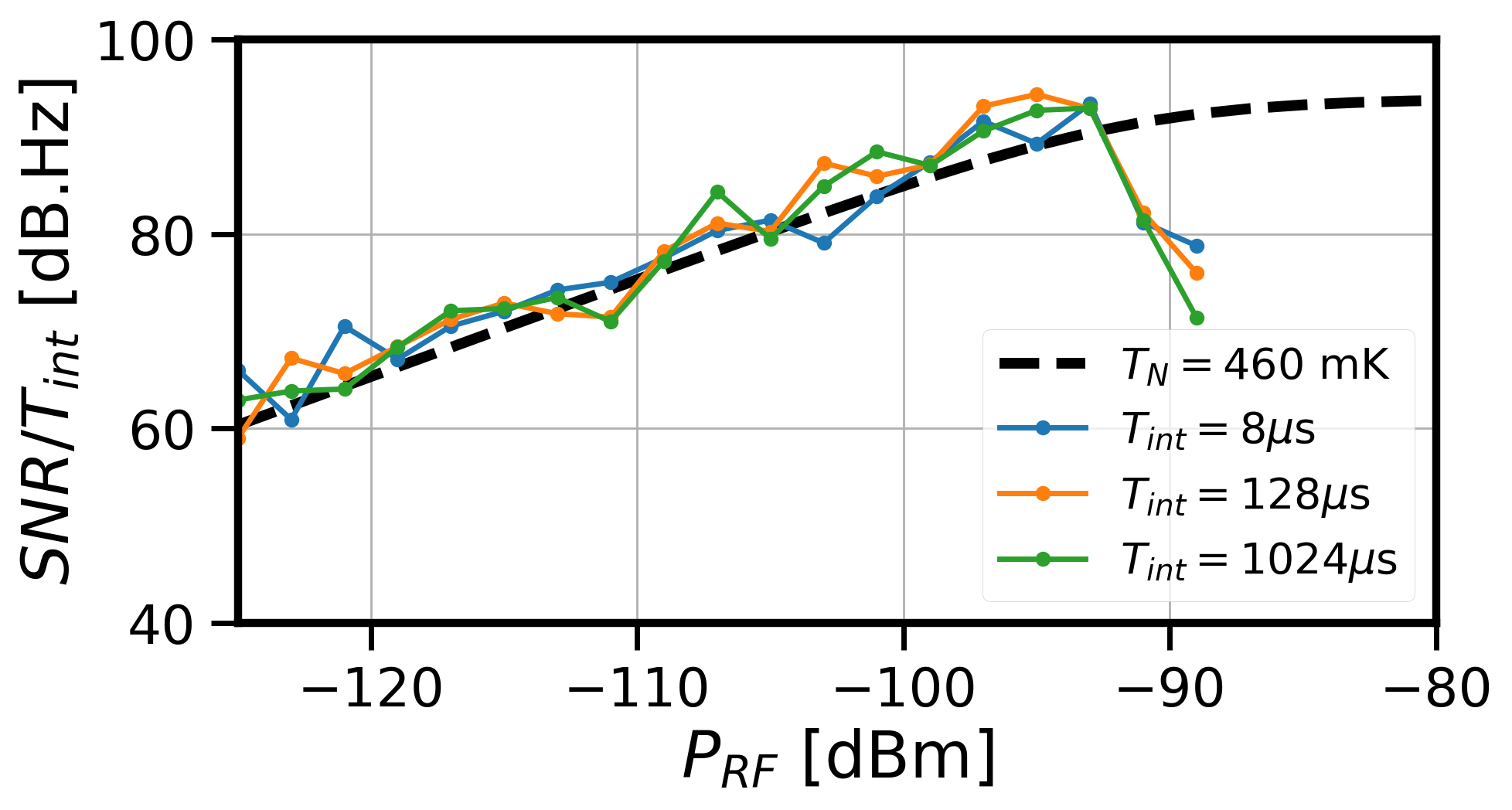} 
    \caption {Measured normalized readout SNR for different $P_{RF}$ and $T_{int}$. The black dotted line indicates the simulated SNR\textsubscript{N}, assuming a $T_{N}$ of 460\,mK.\label{fig:Normalized_SNR}}
\end{figure}

 Fig.\,\ref{fig:Normalized_SNR} shows the measured and simulated SNR\textsubscript{N}. The black dotted line indicates the simulated SNR\textsubscript{N} based on the previously simulated readout signal, with a $T_N$ of 460\,mK. As expected, the data shows that the measured SNR\textsubscript{N} overlaps each other for different $T_{int}$ experiments. A maximum measured SNR\textsubscript{N} of $\sim$89\,dB$\cdot$Hz is observed at $P_{RF}$\,=\,-95\,dBm, and SNR\textsubscript{N} degrades beyond this power level.

\section{Discussion}
The previous section showed that the qubits' resonant frequency shift is power-dependent. This section explores the implications of the power-dependent readout behavior, examines the associated trade-offs, and addresses the noise requirements for the readout chain.

\subsection{Frequency Multiplexing Consideration}
The measurement and model indicate that the readout power can control the qubit's frequency shift. Thus, a higher readout power leads to the possibility of reading out more qubits per given bandwidth, resulting in a higher spectrum efficiency as the frequency spacing requirement between each qubit decreases. As depicted in Fig.\,\ref{fig:freq_shift}, the frequency shift is estimated to be on the order of 2\,MHz at the peak of the SNR\textsubscript{N} with a $P_{RF}$ of -95\,dBm. Assuming each qubit has a frequency shift of 2\,MHz with a 3\,MHz resonator bandwidth, each qubit has a channel bandwidth of $\sim$5\,MHz. However, it should be noted the frequency spacing between the qubits cannot be too small as the readout pulse power, depending on the desired integration times, may leak into neighboring qubits, which eventually may require pulse shaping for a dense FDMA readout architecture.

It is important to note that the number of qubits that can be read out within a specific frequency range can be increased by narrowing the resonator bandwidth through a resonator design with a large characteristic impedance. However, there are practical constraints on how much the resonator bandwidth can be reduced. At very small resonator bandwidth, readout is mostly limited by the ring-up time of the resonator when given an RF impulse. Consequently, the readout response may not settle fully for a given $T_{int}$, thus reducing the readout signal and, eventually, the SNR.

\subsection{Optimum Tunnel Coupling and Readout Operating Frequency}
Based on the discussion above, one could also ask about the preferable readout frequency for gate-based readout. Given a readout frequency $f_r$, one must bias the $t_c$ such that $(2t_c/h)/f_r$\,=\,2. Consequently, the tunnel coupling should be proportionally increased when the qubit is read at a higher $f_r$ to keep the system at the adiabatic limit. A lower frequency shift is expected when the qubit is read out at higher $f_r$ due to the lower quantum capacitance at higher $t_c$ bias. In addition, a higher readout frequency increases the power consumption of the electronics significantly, which can also limit the scalability of the readout system due to the limited power budget in the dilution refrigerator. However, operating at a higher readout frequency may decrease the footprint of the resonator, which can be beneficial when the system is scaled up. 

The opposite behavior is expected when the readout frequency is lowered. Lower readout frequency implies a lower $t_c$. Hence, a larger quantum capacitance is to be expected, which can increase the SNR. As the frequency of operation is lowered, power consumption of the electronics is also expected to follow suit. However, the larger quantum capacitance means that each qubit occupies a larger channel and limits the number of qubits occupying a certain bandwidth. Additionally, pushing the readout frequency lower leads to a larger area required for the resonator and may limit scalability.

 \subsection{Noise Temperature Requirement for Readout Electronics}
 

\begin{figure}
    \centering
    \includegraphics[width=0.45\textwidth]{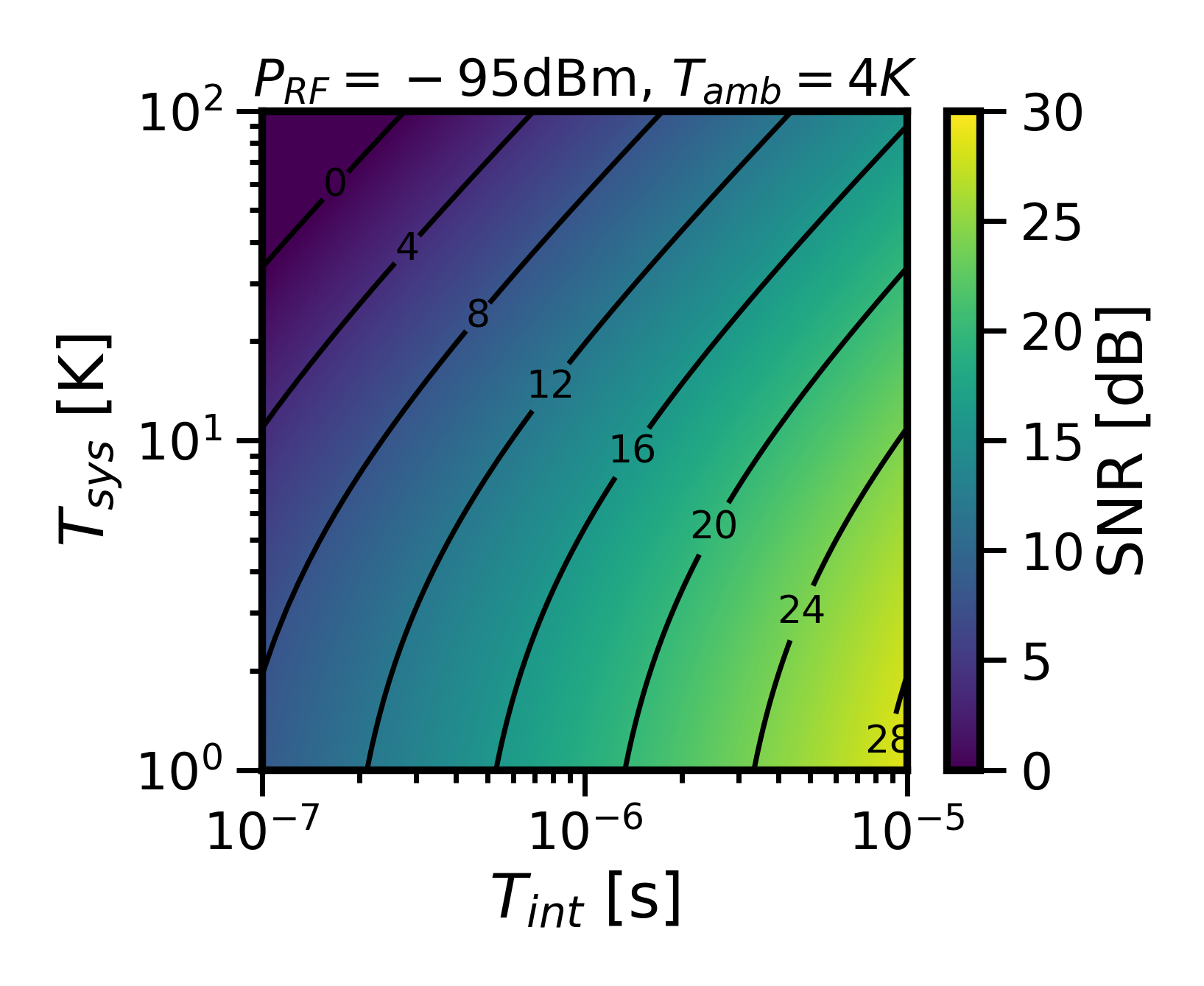} 
    \caption {SNR contour plot for different $T_{int}$ and $T_{sys}$ values with a $T_{amb}$\,=\,4\,K and $P_{RF}$\,=\,-95\,dBm. \label{fig:Tsys_vs_Tint}}
\end{figure}

As discussed in Section I, an SNR larger than 11.5\,dB is needed to achieve BER\,$\leq$\,10\textsuperscript{-4}. Assuming a $T_{int}$ of 1\,$\mu$s, the maximum inherent SNR of the measured gate-based readout is 29\,dB, extrapolated from SNR\textsubscript{N} of 89 dB$\cdot$Hz [see Fig.\,\ref{fig:Normalized_SNR}]. Hence, an excess SNR of 17.5\,dB allows one to increase $N_0$ by 56$\times$ and still satisfy the BER specification, resulting in a maximum $T_{N}$ of 25.8\,K for the readout chain. Furthermore, assuming the readout electronics are now operating at 4\,K, the readout chain should exhibit an input-referred noise temperature ($T_{sys}$) of $\leq$\,21.8\,K to still comply with a readout SNR of 11.5\,dB — a significantly more relaxed noise requirement in comparison to the performance offered by TWPA and the HEMT LNA.

Fig.\,\ref{fig:Tsys_vs_Tint} shows the SNR for different $T\textsubscript{sys}$ and $T_{int}$ values when the system is probed at its maximum achievable $A_{sig}$ ($P_{RF}$\,=\,-95\,dBm) while considering a $T_{amb}$ of 4\,K. Assuming a modest $T_{sys}$ of 50\,K based on the recently published cryo-CMOS RX design \cite{prabowo136to8GHz17mW2021,ruffino13FullyIntegrated40nm2021,parkFullyIntegratedCryoCMOS2021, van_winckelt_28nm_2022, kang_40-nm_2022}, the contour plot reveals that the prior-art cryo-CMOS receivers cannot satisfy the 11.5\,dB target SNR at a $T_{int}$ of 1\,$\mu$s. As it is suspected that the noise performance of the cryo-CMOS receivers is limited by the shot noise \cite{gong_cryo-cmos_2022} and the self-heating effect \cite{t_hart_characterization_2021}, passive amplification techniques should be investigated in the future to avoid large biasing currents required for active devices in LNAs \cite{mehrpoo_cryogenic_2020, prabowoPassive2024}.

\section{Conclusion}
This work presents a theoretical framework that allows one to investigate the impact of the receiver's noise temperature, readout probe power, and integration time on the SNR and fidelity of RF gate-based readout systems. Considering the qubit's quantum mechanical behavior, a semi-classical model was developed that estimates the state- and power-dependent resonance frequency shift and intrinsic readout SNR. Following the trajectory dictated by the developed theory and simulation, the experimental results showed that the frequency spacing between the resonance peaks corresponding to different qubit states was reduced by increasing the power of the readout tone. This affects the separation between the qubit states in the constellation diagram and eventually limits the maximum achievable SNR of the gate-based readout systems. Moreover, based on the outcome of this study, different trade-offs among the qubit biasing, the readout SNR, the system's footprint, the receiver's operating frequency, bandwidth, and noise are discussed. Consequently, the presented guideline can help designers realize a scalable and low-power readout chain while obtaining a readout fidelity sufficient for fault-tolerant quantum computation. 

\section*{Acknowledgment}
This research was funded by Intel Corporation. The authors would like to thank G. Zheng and P. Harvey-Collard for fabricating the device, Z. Y. Chang, J. Mensingh, O. Benningshof, N. Alberts, R. Schouten, R. Vermeulen from TU Delft for measurement support, L. DiCarlo and his team for access to the \textsuperscript{3}He cryogenic measurement setup, L. P. Kouwenhoven and his team for access to the NbTiN film deposition, W. Oliver for providing the TPWA, and the members of the Spin Qubit team, the Cryo-CMOS team, M. F. Gonzalez-Zalba, and S. Pellerano from Intel Corporation for the useful discussions. Data repository in \cite{DOI_data_Bagas}.

\bibliographystyle{IEEEtran}
\bibliography{ref.bib}

\end{document}